\documentclass[11pt]{amsart}

\usepackage[T1]{fontenc}
\usepackage[utf8]{inputenc}

\usepackage[margin=1in]{geometry}
\usepackage{booktabs,tabularx}

\usepackage{csquotes}
\usepackage[USenglish]{babel}

\usepackage{lmodern}
\usepackage{mathtools,amsfonts,amsthm,mathrsfs,amssymb}
\mathtoolsset{showonlyrefs}

\usepackage{textcomp}

\usepackage[
  backend=biber,
  style=numeric,
  sorting=nyt,
  sortcites=true,
  abbreviate=true,
  giveninits=true,
  useprefix=false,
  isbn=false,
  terseinits=true,
  maxbibnames=99,
  maxcitenames=99
]{biblatex}
\usepackage[shortlabels]{enumitem}
\usepackage[dvipsnames]{xcolor}

\PassOptionsToPackage{%
  pdfusetitle,%
  colorlinks=true,%
  pdfstartview=Fit,%
  linkcolor=Blue,%
  citecolor=Green,%
  urlcolor=WildStrawberry,%
  linktoc=page%
}{hyperref}
\usepackage{bookmark}
\usepackage[capitalise,noabbrev]{cleveref}
\usepackage[all]{hypcap} 

\newtheorem{theorem}{Theorem}
\newtheorem{lemma}[theorem]{Lemma}
\newtheorem{proposition}[theorem]{Proposition}
\newtheorem{corollary}[theorem]{Corollary}

\newtheorem{remark}{Remark}

\theoremstyle{definition}

\newcommand*{\ZZ}{\mathbb{Z}}

\newcommand*{\cI}{\mathcal{I}}

\renewcommand{\epsilon}{\varepsilon}

\renewcommand{\Pr}{\mathbb{P}}
\renewcommand{\subset}{\subseteq}
\newcommand{\indicator}[1]{\mathbf{1}_{#1}}

\newcommand{\E}{\mathbb E}

\newcommand{\tv}{d_{\mathrm{TV}}}
\newcommand{\KL}{D_{\mathrm{KL}}}

\newlist{enumerata}{enumerate}{1}
\setlist[enumerata]{label=\upshape{(\alph*)}}
\setlist[enumerate]{label=\upshape{(\roman*)}}

\title[Large independent sets via local occupancy]{Large independent sets via neighborhood block dynamics, Glauber dynamics, and local occupancy}
\date{\today}

\author[E.\ Davies]{Ewan Davies}
\address{Department of Computer Science, Colorado State University, Fort Collins CO, USA}
\email{research@ewandavies.org}

\author[B.\ Tan]{Brian Tan}
\address{Department of Computer Science, Colorado State University, Fort Collins CO, USA}
\email{Brian.Tan@colostate.edu}

\begin{document}

\begin{abstract}
      Local occupancy can be used to bound the expected size of a hard-core independent set.
      The framework of locally stationary distributions shows that for triangle-free graphs, despite slow mixing at the relevant fugacity parameter, standard Glauber dynamics can be used to find sets close to the local occupancy guarantee.

      We improve the application of this framework to the hard-core model by replacing a conditional mixing argument with a more direct approach based on a short sequence of single-site updates that, from every initial configuration, achieves the required occupancy inequality.
      In general graphs, naive conditional mixing estimates or certain problematic structures in the neighborhood can in principle incur running time cost exponential in the maximum degree \(\Delta\).

      We design short update sequences that, via our relaxed criterion, yield running times polynomial in \(\Delta\) for graphs with a hereditary bound \(b\) on the density of \(K_{t,t}\) in neighborhoods.
      Our methods handle the case where \(t\) and \(b\) can grow with the maximum degree, and the resulting algorithm attains the random graph shattering threshold when they grow sufficiently slowly.
      That is, we give improved local occupancy parameters and a short update sequence analysis which together show that Glauber dynamics finds independent sets of size \((1-o(1))n(\log\Delta)/\Delta\) in polynomial time whenever \(t\) and \(b\) are constant or grow suitably slowly.
      In a similar regime, we also establish that the (correspondence) chromatic number is at most \((1+o(1))\Delta/\log\Delta\) which improves upon prior results by a factor of at least \(4\) in the case \(t\ge 4\).
\end{abstract}

\maketitle

\section{Introduction}\label{sec:introduction}

In its elementary form, local occupancy for the hard-core model gives lower bounds on the expected size of an independent set drawn from the hard-core model~\cite{DJPR18,DdKP21,DKPS20a}.
See the recent survey~\cite{DK25} for a discussion of the history of the method and its applications to independent sets and graph coloring.
In recent work, \citeauthor{LMR+24} introduced the \emph{locally stationary distributions}~\cite{LMR+24,LMR+25} as a framework for analyzing slow-mixing Markov chains in the context of optimization.
One of their motivating examples is the hard-core model on triangle-free graphs, and they show that standard Glauber dynamics gives polynomial-time constructions independent sets of size close to the equilibrium guarantee offered by local occupancy, despite slow mixing.
Local occupancy parameters are, however, available in a wide range of settings beyond triangle-free graphs.
In many cases the resulting lower bounds on the independence number are asymptotically the best known, but in graphs with triangles the method does not immediately yield a polynomial-time algorithm for finding independent sets of the guaranteed size.
We investigate the details of what conditions beyond local occupancy give rise to algorithmic constructions via Glauber dynamics or related Markov chains, and extend the framework of locally stationary distributions subject to the usual local occupancy condition.
The core question is when a suitable local dynamics can, in polynomial time, find independent sets of size close to the equilibrium guarantee offered by local occupancy.
Our main theoretical contribution is an analysis for Glauber dynamics based on short sequences of single-site updates in a neighborhood: we show that it suffices that these updates achieve the local occupancy inequality from every initial configuration.
This replaces a conditional mixing argument in the original framework~\cite{LMR+24,LMR+25} and can lead to improved running times.

Our main application concerns graphs with a hereditary bound on the density of copies of \(K_{t,t}\) in their neighborhoods.
For simplicity, we state the bounds for fixed \(t\) and allow the hereditary density bound \(b=b(\Delta)\) to grow with the maximum degree.
\cref{rem:ktt-growing-t} gives the extension to growing \(t\) which follows from the same analysis.
The setting includes triangle-free graphs (\(b=0\) and \(t=1\)), graphs with bounded maximum average degree in neighborhoods (\(b\ge0\) and \(t=1\)), and \(K_{1,t,t}\)-free graphs (\(b=0\)).
With maximum degree \(\Delta\) and the parameter \(t\) (fixed, for simplicity) and the density \(b=b(\Delta)\), let \(\rho=\Delta/(b+1)^{1/(2t-1)}\).
By developing improved local occupancy parameters we show that ordinary Glauber dynamics finds independent sets of size \((1-o(1))n\log\rho/\Delta\) in polynomial time whenever \(\rho\to\infty\).
The estimates are uniform in \(b\), so they apply throughout the range where \(b+1=o(\Delta^{2t-1})\).
For any \(b+1=\Delta^{o(1)}\), this gives \((1-o(1))n\log\Delta/\Delta\), asymptotically attaining the random-graph shattering benchmark.
For fixed \(b\), this matches the leading constant of the guarantee of one of the main results in~\cite{DJM25}; we discuss the comparison in detail below.

We also give a simple observation that a specific neighborhood block dynamics achieves the local occupancy guarantee, which gives a much simpler analysis of an algorithm whose running time is polynomial in \(n\) and (potentially) exponential in the maximum degree than the local stationarity framework for Glauber dynamics.
This approach was hinted at, but not explored in prior work~\cite[Rem.~IV.5]{LMR+24}.
As our version of neighborhood block dynamics corresponds precisely to the local occupancy inequality, we give a complete analysis as an important stepping-stone to the Glauber dynamics result.

Our fixed update sequence analysis is important in two ways. First, it avoids the need to show conditional mixing. This can be burdensome when it does hold, but more importantly our condition is weaker as, e.g., large balanced bipartite graphs in neighborhoods obstruct conditional mixing but not our criterion. 
Second, the proofs establishing local occupancy parameters hint at how to construct suitable update sequences. Despite having no formal relationship to Glauber dynamics and update sequences, the analysis of the balance between $1/Z_F(\lambda)$ and $E_F(\lambda)$ inherent to local occupancy in triangle-free graphs, graphs with few triangles, and our bounded $K_{t,t}$ density in neighborhoods setting guides the design of the sequences. See \cref{sec:sweeps,sec:ktt-sweeps}.

The extension of the local stationarity framework to avoid local conditional mixing is important in the main application.
An argument requiring uniform mixing of the full conditional neighborhood chain at our chosen fugacity cannot provide the same guarantees, even for \(t=2\).
In \cref{prop:slow-neighborhood} we construct \(K_{1,2,2}\)-free graphs whose conditional neighborhood chains do not exhibit rapid mixing.
The obstruction holds for every polynomial-length prescribed sequence of single-site updates, although three sweeps satisfy our averaged occupancy certificate.
The intuition behind the construction is simply that, despite \(K_{1,2,2}\)-freeness, a neighborhood can contain a large balanced bipartite graph, which is known to be a slow-mixing example for Glauber dynamics at the fugacity we need to use for the best asymptotic bounds.
But bipartite graphs are not an obstruction to our (averaged) short update sequence analysis.

We also strengthen our novel local occupancy analysis to the setting of \emph{chromatic} local occupancy (see~\cite{DKPS20a}) and obtain, when \(\log\rho =\omega(t\log\log\Delta)\) (this holds, e.g.\ for fixed \(t,b\)), the correspondence coloring bound \[ \chi_c(G)\leq(1+o(1))\frac{\Delta}{\log\rho}.
\]
In particular, \(b+1=\Delta^{o(1)}\) gives leading constant one in front of \(\Delta/\log\Delta\).
For fixed \(t,b\), earlier bounds had leading constants \(t\) from local occupancy~\cite{DKPS20a} and \(4\) from the nibble method~\cite{ABD25}.
Provided we carefully convert between the hereditary and non-hereditary settings, our result also improves the leading constant of results of \citeauthor{ADK26}~\cite{ADK26}.
For \(K_{1,t,t}\)-free graphs, our leading constant one holds throughout the range \(t=o(\log\Delta/\log\log\Delta)\), the same range in which~\cite{ABD25} gives leading constant four.
The nibble method results~\cite{ABD25,ADK26} are substantially stronger in the important sense that they additionally apply in the color-degree setting.
We give the details of coloring parameters, proofs, and precise comparisons in \cref{app:chromatic-occupancy}.

\subsection{Notation and main results}\label{subsec:notation}

Let \(G=(V,E)\) be a finite simple graph on \(n\) vertices with maximum degree at most \(\Delta\).
Write \(N(u)\) for the open neighborhood of \(u\) and \(\cI(G)\) for the independent sets of \(G\), including the empty set.
For a graph \(G\) and fugacity \(\lambda>0\), set
\begin{align}
      Z_G(\lambda) & =\sum_{I\in\cI(G)}\lambda^{|I|}, & E_G(\lambda) & =\frac{\lambda Z_G'(\lambda)}{Z_G(\lambda)}, & p=\frac{\lambda}{1+\lambda}.
\end{align}
Since we will consider these quantities for a graph \(\emptyset\) with no vertices and
no edges, note for clarity that
\begin{align*}
      Z_{\emptyset}(\lambda) & =1, & E_{\emptyset}(\lambda) & =0.
\end{align*}
We identify an independent set with its spin configuration, assigning spin \(1\) to occupied vertices and \(0\) to unoccupied vertices.
A configuration is feasible if no adjacent vertices both have spin \(1\).
The hard-core measure on \(\cI(G)\) is given by
\begin{equation}\label{eq:intro-hardcore}
      \mu_{G,\lambda}(I)=\frac{\lambda^{|I|}}{Z_G(\lambda)}.
\end{equation}
With a graph \(G\) and fugacity \(\lambda>0\) as fixed context, we write \(\pi\) for the hard-core measure \(\mu_{G,\lambda}\) and \(\E_\pi\) for expectation with respect to \(\pi\).
Following~\cite{DKPS20a,DK25}, we say that \(G\) satisfies \emph{local occupancy} with parameters
\((\lambda,\beta,\gamma)\), where \(\beta,\gamma>0\), if, for every vertex \(u\in V\),
every induced subgraph \(F\) of \(G[N(u)]\), including the empty graph \(\emptyset\), satisfies
\begin{equation}\label{eq:intro-occupancy}
      \beta\frac{p}{Z_F(\lambda)}+\gamma E_F(\lambda)\geq1.
\end{equation}
Throughout, we write \(\Gamma=\beta+\gamma\Delta\) for brevity.
An elementary computation based on conditioning on the exterior of each neighborhood and summing over the vertices gives the equilibrium bound \(\E_\pi|I|\geq n/\Gamma\); see e.g.~\cite{DJPR18,DdKP21,DK25}.

Let \(P_u\) be the heat-bath kernel that updates \(u\): if no neighbor is occupied, it sets \(u\) to be occupied with probability \(p\) and unoccupied otherwise; if a neighbor is occupied, it sets \(u\) to be unoccupied.
Let \(Q_u\) be the kernel for the following update: resample \(N(u)\) from its conditional probability distribution under \(\pi\), keeping \(V\setminus N(u)\) fixed.
Distributions multiply kernels on the left, so \(Q_uP_u\) first resamples \(N(u)\) and then updates \(u\).
The kernels of ordinary random-scan Glauber dynamics and the specific neighborhood block dynamics that we study are, respectively,
\begin{align}\label{eq:dynamics-kernels}
      K^G & =\frac1n\sum_{u\in V}P_u, & K^N & =\frac1n\sum_{u\in V}Q_uP_u.
\end{align}
For an initial probability distribution \(\nu_0\) on \(\cI(G)\), define \(H_0=\KL(\nu_0\Vert\pi)\) where \(\KL\) denotes relative entropy.
For \(\lambda>0\) let \[ H_\lambda=n\log\frac{1+\lambda}{\min\{1,\lambda\}}, \] and note that we always have \(H_0\leq H_\lambda\).
Indeed, \(Z_G(\lambda)\leq(1+\lambda)^n\) and \(\lambda^{|I|}\geq\min\{1,\lambda\}^n\), so \(\pi(I)\geq e^{-H_\lambda}\) for every independent set \(I\), and hence \(\KL(\nu_0\Vert\pi)\leq H_\lambda\).
For either dynamics, let \(I_j\) denote the state after \(j\) steps and for an integer time bound \(T\ge 1\) let \(J\) be independent and uniform on \(\{0,\ldots,T-1\}\).
Throughout the statements below, \(\epsilon\in(0,1)\) is an accuracy parameter.

Our first result is essentially a warm-up which shows that neighborhood block dynamics achieves the local occupancy guarantee in expected size, with a simple entropy-drop argument.
In general, the running time of implementing neighborhood block dynamics can be exponential in the maximum degree \(\Delta\), as we do not have general-purpose algorithms for sampling from the conditional hard-core measure on a neighborhood time polynomial in \(\Delta\).

\begin{theorem}[Neighborhood block dynamics]\label{thm:block}
      Suppose \(G\) satisfies local occupancy with parameters \((\lambda,\beta,\gamma)\).
      From any initial probability distribution \(\nu_0\) and for any time bound \(T\ge1\), with \(J\) uniform on \(\{0,1,.
      \dotsc,T-1\}\) neighborhood block dynamics satisfies
      \begin{equation}\label{eq:block-bound}
            \E|I_J|\geq\frac{n}{\Gamma}
            \left(1-\sqrt{\frac{(\beta^2 + \gamma^2\Delta^2)H_0}{2T}}\right).
      \end{equation}
      In particular, for \(0<\epsilon<1\) and \[ T\geq\frac{(\beta^2 + \gamma^2\Delta^2)H_\lambda}{2\epsilon^2}, \] we have \(\E|I_J|\geq(1-\epsilon)n/\Gamma\).
      Each block update can be implemented in time \(2^{O(\Delta)}\), for example by enumerating independent sets in the neighborhood.
\end{theorem}

To obtain a guarantee for ordinary Glauber dynamics, we require a condition that a fixed sequence of single-site updates of the neighbors approximately achieves the local occupancy inequality from every configuration.
The fixed sequence is a feature of the analysis; the algorithm is standard random-update Glauber dynamics.
For each \(u\), consider a word \((v_{u,1},\ldots,v_{u,\ell_u})\) of vertices in \(N(u)\), allowing repetitions, and set
\begin{align}
      S_u & =P_{v_{u,1}}\cdots P_{v_{u,\ell_u}}, & \kappa & =\max_{v\in V}\sum_{u\in V} |\{i:v_{u,i}=v\}|.
\end{align}
These words are fixed for the purpose of the analysis, and are independent of the initial configuration and of all random choices made by the algorithm.
Thus \(\kappa\) is the maximum, over vertices \(v\), of the total number of occurrences of \(v\) across all the words.
As an example, if each word corresponds to a single full sweep of \(N(u)\) then a vertex \(v\) appears once for each \(u\in N(v)\) and \(\kappa=\max_{v\in V}d_G(v)\leq\Delta\).

\begin{theorem}[Glauber dynamics]\label{thm:transfer}
      Let \(\lambda,\beta,\gamma>0\), \(0\leq\eta<1\), and for each \(u\) consider a fixed-word neighborhood update kernel \(S_u\) as above.
      For this collection of words, consider \(\kappa\) as specified above.
      For an \(n\)-vertex graph \(G\) with maximum degree \(\Delta\), suppose that for every probability distribution \(\nu\) on \(\cI(G)\) and every vertex \(u\),
      \begin{equation}\label{eq:word-certificate}
            \beta\Pr_{\nu S_uP_u}(u\in I)
            +\gamma\E_{\nu S_u}|I\cap N(u)|\geq1-\eta.
      \end{equation}
      With \(p=\lambda/(1+\lambda)\), set \(C=\beta+\kappa\max\{\beta p,\gamma\Delta\}\).
      For any initial probability distribution \(\nu_0\), and any time bound \(T\ge1\), with \(J\) uniform on \(\{0,1,\ldots,T-1\}\) ordinary Glauber dynamics satisfies
      \begin{equation}\label{eq:glauber-bound}
            \E|I_J|\geq\frac{n}{\Gamma}
            \left(1-\eta-C\sqrt{\frac{H_0}{2T}}\right).
      \end{equation}
      In particular, \(T\geq C^2H_\lambda/(2\epsilon^2)\) gives expected size at least \((1-\eta-\epsilon)n/\Gamma\).

      The same conclusion holds if \eqref{eq:word-certificate} is replaced
      by an average over fixed prefixes \(S_{u,1},\ldots,S_{u,r_u}\) of
      \(S_u\), with \(r_u\geq1\):
      \begin{equation}\label{eq:averaged-word-certificate}
            \frac1{r_u}\sum_{j=1}^{r_u}
            \left(\beta\Pr_{\nu S_{u,j}
                  P_u}(u\in I) +\gamma\E_{\nu S_{u,j}}|I\cap N(u)|\right)\geq1-\eta.
      \end{equation}
      Here, \(\kappa\) still counts occurrences in the full words \(S_u\).
\end{theorem}

Note that no separate equilibrium local occupancy assumption is required; the update-sequence certificate is the sole local occupancy input.
This omission of the traditional local occupancy hypothesis is mild, however, by considering \(\nu=\pi\) the update-sequence hypothesis directly implies the equilibrium bound \(\E_\pi|I|\geq (1-\eta)n/\Gamma\) analogous to local occupancy.
Note that it suffices to verify the update sequence certificate inequality (i.e.\ condition~\eqref{eq:word-certificate} or~\eqref{eq:averaged-word-certificate}) from each initial configuration, since expectations are linear in the input probability distribution.

The local stationarity analysis we give for \cref{thm:transfer} retains the entropy accounting of the original local stationarity analysis~\cite{LMR+25}, while replacing the conditional mixing comparison by the local occupancy inequalities in \cref{thm:transfer}.
The main idea interprets the local occupancy condition for a vertex as its ``score'' and shows that if a probability distribution is approximately unchanged by each single-site update, a simple telescoping sum shows that its score is close to the score after that update sequence.
Entropy analysis then controls the accumulated error along the ordinary dynamics.
This requires only the occupancy inequality we assume for \cref{thm:transfer}, which can hold well before the conditional probability distribution on the neighborhood is close to equilibrium.

At a high level this approach resembles a recursive application of the original observation due to Liu et al.~\cite{LMR+24} that certain score bounds can hold long before mixing, applied inside the neighborhoods of the input graph.
In practice, we do not strictly recursively apply the original observation inside neighborhoods, but rather we exploit some slack in known analyses of local occupancy, well-illustrated in the case of graphs with bounded maximum average degree in neighborhoods (see \cref{thm:sweeps} and its proof), to establish the stronger update-sequence criterion.
That is, the very proof of traditional local occupancy inequalities inspires the belief that a short sequence of single-site updates can achieve the occupancy inequality from every initial configuration, and our algorithmic analysis shows that this is indeed sufficient to guarantee large independent sets in polynomial time.

\Cref{prop:slow-neighborhood} gives a strong separation of the short update sequence hypothesis and the conditional mixing hypothesis (e.g. represented by a modified log-Sobolev constant) at a realistic fugacity \(\lambda=(\log\Delta)^{-3}\): on a family of \(K_{1,2,2}\)-free graphs, the three-sweep averaged certificate holds from every initial configuration, but no polynomial-length prescribed update sequence uniformly approximates a conditional neighborhood distribution even to total variation error \(1/4\).
This obstruction forces a superpolynomially small conditional modified log-Sobolev constant.

For our main application, fix an integer \(t\geq1\).
For a graph \(F\), let \(N_t(F)\) count copies of \(K_{t,t}\) in \(F\), i.e.\ unordered pairs of disjoint \(t\)-sets in \(V(F)\) with every edge between the two sets present.
Edges within the sets are permitted, so these copies of \(K_{t,t}\) need not be induced.
For \(b\geq0\), we say that \(F\) has \emph{\(K_{t,t}\)-density at most \(b\)} if
\begin{equation}\label{eq:ktt-hereditary-budget}
      2tN_t(F)\leq b|V(F)|.
\end{equation}
Since each copy of \(K_{t,t}\) contains \(2t\) vertices, this condition says that in expectation, a uniform random vertex of \(F\) is contained in at most \(b\) copies of \(K_{t,t}\).
For \(t=1\) the condition is that \(F\) has average degree at most \(b\).
We say that \(G\) has \emph{hereditary \(K_{t,t}\)-density at most \(b\) in neighborhoods} if every induced subgraph of every neighborhood subgraph \(G[N(u)]\) has \(K_{t,t}\)-density at most \(b\).

\begin{theorem}\label{thm:ktt-occupancy}
      For \(\Delta\ge 1\), \(t\geq1\), and \(b\geq0\), suppose that \(G\) has maximum degree \(\Delta\) and hereditary \(K_{t,t}\)-density at most \(b\) in neighborhoods.
      Write \(\rho=\Delta/(b+1)^{1/(2t-1)}\).
      For sufficiently large \(\rho\), with a threshold depending only on \(t\), at fugacity \[ \lambda=\frac1{(b+1)^{1/(2t-1)}(\log\rho)^{2t-1}}, \] there are parameters \(\beta,\gamma>0\) with \[ \beta+\gamma\Delta =\left(1+O_t \left(\frac{\log\log\rho}{\log\rho}\right)\right) \frac{\Delta}{\log\rho} \] such that \(G\) satisfies local occupancy with parameters \((\lambda,\beta,\gamma)\).
      The estimate is uniform in \(b\) as \(\rho\to\infty\).
\end{theorem}

Previous local occupancy analysis for \(K_{1,t,t}\)-free graphs~\cite{DKPS20a} gave (for fixed \(t\)) parameters with \(\Gamma\sim t\Delta/\log\Delta\).
That proof simply uses the Kővári--Sós-Turán theorem~\cite{KST54} to bound the maximum average degree of a neighborhood, and then applies an analysis tailored to this setting.
Here, we obtain an improvement by exposing the proof method of~\cite{KST54} to the local occupancy analysis, which allows us to exploit the full hereditary \(K_{t,t}\)-density bound in neighborhoods.
See \cref{lem:ktt-weighted}, and note that the proof is a light adaptation of the original to handle weights coming from marginals in the hard-core model.

Our main application follows.
For completeness we state that an algorithm finds a large independent set with high probability, obtaining this from the expected-size guarantee offered by \cref{thm:transfer} is entirely standard.

\begin{corollary}\label{cor:ktt-glauber}
      For \(\Delta\ge 1\), \(t\geq1\), and \(b\geq0\), suppose that \(G\) has maximum degree \(\Delta\) and hereditary \(K_{t,t}\)-density at most \(b\) in neighborhoods.

      For each vertex \(u\in V(G)\), let \(S_u\) consist of three full sweeps of \(N(u)\), each in the same arbitrary fixed order, and let \(S_{u,1}\) be the prefix corresponding to two sweeps and \(S_{u,2}=S_u\).
      For sufficiently large \(\rho\), with a threshold depending only on \(t\), with the parameters \((\lambda,\beta,\gamma)\) from \cref{thm:ktt-occupancy} (made explicit in \cref{prop:ktt-occupancy}), condition~\eqref{eq:averaged-word-certificate} of \cref{thm:transfer} holds with \(\eta=0\) and \(\kappa\leq3\Delta\).

      Writing \(p=\lambda/(1+\lambda)\), together with Theorem~\ref{thm:transfer} this gives that from every initial state and time bound \[ T\geq\frac{(\beta+3\Delta\max\{\beta p,\gamma\Delta\})^2H_\lambda}{2\epsilon^2}, \] with \(J\) uniform on \(\{0,1,\dotsc,T-1\}\), ordinary Glauber dynamics satisfies \(\E|I_J|\geq(1-\epsilon)n/\Gamma\).

      Retaining the largest visited set and repeating independently finds an
      independent set of size
      \begin{equation}\label{eq:intro-density-guarantee}
            \left(1-O_t\!\left(\frac{\log\log\rho}{\log\rho}\right)\right)
            \frac{n\log\rho}{\Delta}
      \end{equation}
      with probability at least \(1-\delta\) in time polynomial in \(n\) and
      \(\log(1/\delta)\), for \(0<\delta<1\).
      The asymptotic estimate is uniform in \(b\) as \(\rho\to\infty\).

      In particular, if \(\log(b+1)=o(\log\Delta)\), then as \(\Delta\to\infty\) we have a polynomial-time algorithm which finds an independent set of size \((1-o(1))n\log\Delta/\Delta\) with high probability.
\end{corollary}

\begin{remark}\label{rem:ktt-growing-t}
      The same parameters and three-sweep certificate allow \(t\) and \(b\) to grow simultaneously.
      Write \(D=\log\rho\), and suppose that \(D\to\infty\) and \(t\log D=o(D)\).
      Then \cref{thm:ktt-occupancy,cor:ktt-glauber} hold with their relative error terms replaced by \(O(t\log D/D)\), with an absolute implicit constant.
      In particular, ordinary Glauber dynamics finds an independent set of size \[ \left(1-O\!
            \left(\frac{t\log D}{D}\right)\right)
            \frac{nD}{\Delta}
      \]
      with high probability in polynomial time, with a polynomial exponent independent of \(t\).
      When \(b=0\), this gives \((1-o(1))n\log\Delta/\Delta\) for \(K_{1,t,t}\)-free graphs throughout \(t=o(\log\Delta/\log\log\Delta)\).
      The necessary estimates for these conclusions are included in the proofs of \cref{prop:ktt-occupancy,cor:ktt-glauber}.
\end{remark}

When \(t=1\), the hypothesis says exactly that every neighborhood has maximum average degree at most \(b\), since \(N_1(F)=|E(F)|\).
More generally, all our expected-size bounds can be converted to high-probability constructions by retaining the largest visited set and repeating independently, as detailed in \cref{rem:amplification}.

We give the details of graph coloring consequences of the ideas behind \cref{thm:ktt-occupancy} (including the required slight strengthening of the conclusion) in \cref{app:chromatic-occupancy}.

\subsection{Prior work and motivation}

A greedy algorithm always finds an independent set of size at least \(n/(\Delta+1)\) in an \(n\)-vertex graph with maximum degree \(\Delta\).
Restrictions on local graph structure can improve this benchmark substantially, and both existence and algorithmic results in this context are well-studied in extremal and probabilistic combinatorics.
In triangle-free graphs, \citeauthor{AKS80}~\cite{AKS80} obtained the logarithmic improvement \(\Omega(n\log\Delta/\Delta)\), and Shearer~\cite{She83} proved the bound \((1-o(1))n\log\Delta/\Delta\) (these results notably also hold in the more general setting of given average degree, though for comparison with local occupancy we retain a focus on maximum degree).
Various results for other local structure follow from the analysis of~\cite{AKS80}, but with some degradation in the leading constant of the bounds.
Shearer's ideas have recently been extended in several directions and permit both stronger conclusions in terms of (local) fractional coloring and weaker hypotheses in terms of more general local structure.
The first improvement is in the local fractional coloring setting introduced by \citeauthor{KP24}~\cite{KP24}, to which Shearer's method for triangle-free was extended by \citeauthor{MS25}~\cite{MS25}.
Shearer's method has been adapted to the presence of triangles~\cite{LRZ01,LRZ02}, and in more recent works the fractional setting was considered in the presence of triangles~\cite{Dav26} and \(r\)-colorable neighborhoods~\cite{Dav26,Dha26a}.
Applications of these ideas in the setting of given degeneracy also form an active area of research~\cite{MS25,Dha26a,ADN26,BFS+26}.

The equilibrium hard-core bounds and local occupancy methods of~\cite{DJPR18,DdKP21,DKPS20a} provide a general way to obtain similar guarantees with good leading constants from more varied local structure.
The method of \citeauthor{AKS80}~\cite{AKS80} and Shearer's induction~\cite{She83} naturally correspond to algorithmic constructions, but it was not until the work of \citeauthor{LMR+24}~\cite{LMR+24,LMR+25} that an algorithmic principle which directly finds large independent sets based on local occupancy was established.
This was a striking omission, as the algorithmic applications of the nibble and of local occupancy (and related ideas) to graph coloring have been studied for some time~\cite{Joh96,Joh96a,AKS99,Mol19,DKPS20a,DKPS20b}.
Of course, a coloring algorithm can be used to find large independent sets, but one should perhaps hope to avoid the complexities of coloring for the sake of an algorithm that constructs large independent sets.
Our simple neighborhood block dynamics observation fills this gap in the sense that it provides a natural algorithm for independent sets tailored to the local occupancy condition, and the short update sequence argument elegantly extends this to ordinary Glauber dynamics and can avoid incurring running time costs exponential in the maximum degree \(\Delta\).

The coloring framework of~\cite{DKPS20a} requires a stronger \emph{chromatic} local occupancy condition, which controls a lower quantile rather than only an expectation (in the term multiplied by \(\gamma\) in~\eqref{eq:intro-occupancy}).
In \cref{app:chromatic-occupancy}, we establish this condition and deduce a novel correspondence coloring bound.

The asymptotic size \(n\log\Delta/\Delta\) achieved in \cref{cor:ktt-glauber} when \(b+1=\Delta^{o(1)}\) corresponds to the familiar algorithmic benchmark associated with \emph{shattering} in sparse random graphs~\cite{AC08,CE15}.
For independent sets in sparse random graphs of average degree \(d\), shattering occurs just above the scale \(n\log d/d\): independent sets of a given size split into exponentially many clusters, each containing an exponentially small fraction of the sets, with distinct clusters separated by symmetric difference \(\Omega(n)\)~\cite{CE15}.
The same scale governs local algorithms on random \(\Delta\)-regular graphs: Rahman and Vir\'ag~\cite{RV17} showed that the largest asymptotic density they can achieve is \((1+o_\Delta(1))\log\Delta/\Delta\), half the maximum independent-set density to leading order.
This motivates \(n\log\Delta/\Delta\) as an algorithmic benchmark for locally sparse graphs.

\citeauthor{DJM25}~\cite{DJM25} give a nibble-based construction of independent sets of size \((1-o(1))n\log d/d\) in \(K_{T,T,T}\)-free graphs, for fixed \(T\), where \(d\) is the average degree.
For fixed \(t,b\), their theorem applies to our graph class: put \(T=\lceil t(b+1)^{1/(2t-1)}\rceil\).
A copy of \(K_{1,T,T}\) would give a neighborhood subgraph on \(2T\) vertices with \(K_{t,t}\)-density at least \[ \frac{t}{T}\binom{T}{t}^2\geq(T/t)^{2t-1}>b, \] so our hypothesis implies \(K_{1,T,T}\)-freeness and hence \(K_{T,T,T}\)-freeness.
Their guarantee in terms of the \emph{average} degree is stronger.
Vertex deletion ideas in the cleaning step of~\cite{DJM25} may permit a transfer of our bounds which require a given maximum degree \(\Delta\) to the case of a given average degree \(d\) without incurring a loss in the leading constant, but we do not pursue this here.

Conversely, \(K_{T,T,T}\)-freeness does not bound hereditary neighborhood density: for \(T\geq2\), \(K_{1,m,m}\) is \(K_{T,T,T}\)-free, while its apex neighborhood contains \(\binom{m}{t}^2\) copies of \(K_{t,t}\).
It would be interesting to extend our techniques to \(K_{T,T,T}\)-free graphs, but we expect this to require additional ideas.

A conjecture of Alon, Krivelevich and Sudakov~\cite{AKS99} states that for every fixed \(F\) there is a constant \(c_F>0\) such that \(\chi(G)\le (c_F+o(1))\Delta/\log\Delta\) holds for all \(F\)-free graphs of maximum degree \(\Delta\), as \(\Delta\to\infty\).
Here, we show that one may take \(c_F=1\) for all \(F\subset K_{1,t,t}\) (and all \(t\ge 1\)).
Recently, OpenAI claimed a full resolution of the conjecture~\cite{OpenAI-AKS} and thus there is perhaps increased importance of determining optimal constants \(c_F\).
For no graph \(F\) containing a cycle do we have a proof that \(c_F < 1\), but a standard random graph construction due to Bollob\'as~\cite{Bol81} shows that for any graph \(F\) containing a cycle we have \(c_F\ge1/2\).
As this work is primarily concerned with the algorithmic construction of large independent sets, we defer to prior works~\cite{ABD23,ABD25,ADK26} for a thorough discussion of the history of the conjecture, important results and known bounds on \(c_F\).

\subsection{Organization}

In \cref{sec:setup} we derive conditional resampling identities which relate local occupancy to heat-bath dynamics, and also establish the local occupancy improvements represented by \cref{thm:ktt-occupancy}.
In \cref{sec:block} we give a proof of the neighborhood block dynamics result, \cref{thm:block}, using a simple identity for the drop of entropy as the algorithm progresses, and in \cref{sec:transfer} we give the additional analysis which, via a certificate that a short sequence of updates approximately achieves the local occupancy inequality in expectation, yields \cref{thm:transfer}.
For our main application to Glauber dynamics, in \cref{sec:sweeps} we verify simple one- and two-sweep conditions which suffice for graphs with no triangles and with bounded maximum average degree in neighborhoods.
This serves as a warm-up for the more general hereditary \(K_{t,t}\)-density case.
In \cref{sec:ktt-sweeps} we prove \cref{cor:ktt-glauber} using three sweeps and an analysis of the average local occupancy score after two and three sweeps are completed.
Finally, in \cref{sec:mixing-comparison} we demonstrate that the short update sequence hypothesis is weaker than conditional mixing in a natural context relevant to our main application.
In \cref{app:chromatic-occupancy} we establish chromatic local occupancy for graphs with a hereditary \(K_{t,t}\) density bound in neighborhoods, and its consequence for correspondence coloring.

\section{Local occupancy and conditional resampling}
\label{sec:setup}

Fix a feasible spin configuration corresponding to an independent set \(J\) on \(V\setminus N(u)\).
A vertex of \(N(u)\) is \emph{available} if it has no occupied neighbor in \(V\setminus N(u)\) (i.e.\ no neighbor in \(J\)).
Let \(F=F(u,J)\) be the subgraph induced by these available vertices.
All other neighbors are forced to be unoccupied, and the conditional probability distribution on \(F\) is the hard-core model on \(F\) at fugacity \(\lambda\).
Notice that \(u\) is part of the fixed exterior in our analysis, which is in line with much prior work on local occupancy.
One can instead consider \(u\) as part of the neighborhood, as in~\cite{LMR+24,LMR+25}, but this complicates the transfer of established local occupancy inequalities.

Conditional on a fixed independent set \(J\subset V\setminus N(u)\), let \(J'\) be obtained from \(J\) by applying \(Q_u\) and then \(J''\) obtained by subsequently applying \(P_u\).
Then elementary conditional probability calculations give
\begin{align*}
      \Pr(u\in J'')   & = \frac{p}{Z_F(\lambda)},       \\
      \E|N(u)\cap J'| & =\E|N(u)\cap J''|=E_F(\lambda).
\end{align*}
Consequently, for every input probability distribution \(\nu\) on \(\cI(G)\), local occupancy
gives (where \(I\) is the random independent set drawn from the measures indicated with subscripts)
\begin{equation}\label{eq:block-certificate}
      \beta\Pr_{\nu Q_uP_u}(u\in I)+\gamma\E_{\nu Q_uP_u}|I\cap N(u)|=\beta\Pr_{\nu Q_uP_u}(u\in I)+\gamma\E_{\nu Q_u}|I\cap N(u)|\geq1.
\end{equation}
Taking \(\nu\) to be the hard-core measure \(\pi\) (which is stationary for the kernels \(P_u\) and \(Q_u\)), summing over \(u\), and using \(\sum_u |I\cap N(u)|=\sum_{v\in I}d_G(v)\leq\Delta|I|\) yields the equilibrium guarantee \(\E_\pi|I|\geq n /\Gamma\) originally proved in~\cite{DJPR18} for triangle-free graphs, and subsequently extended to more general settings~\cite{DdKP21,DKPS20a}.

\subsection{Hereditary \texorpdfstring{\(K_{t,t}\)}{Ktt} density in neighborhoods}
\label{subsec:ktt-occupancy}

Recall that \(N_t(F)\) counts copies of \(K_{t,t}\) as subgraphs, with unordered parts.
The following extension of the common neighbor counting argument behind the K\H{o}v\'ari--S\'os--Tur\'an Theorem~\cite{KST54} is used in both the local occupancy analysis and the neighborhood sweep analysis.

\begin{lemma}
      \label{lem:ktt-weighted}
      Let \(t\geq1\) be an integer.
      Let \(b\geq0\) and suppose that the graph \(F\) has \(k\geq1\) vertices and \(K_{t,t}\)-density at most \(b\).
      Set \(s=(b+1)^{1/(2t-1)}\).
      For weights \(0\leq z_v\leq p\) on the vertices of \(F\), we have
      \begin{equation}\label{eq:ktt-weighted}
            \sum_{v\in V(F)}d_F(v)z_v
            \leq 4tspk+4ts^{1/t}k^{1-1/t}\sum_{v\in V(F)}z_v.
      \end{equation}
      Moreover, the average degree \(d\) of \(F\) satisfies
      \begin{equation}\label{eq:ktt-degree}
            d+1\leq 4ts\max\{1,(k/s)^{1-1/t}\}.
      \end{equation}
\end{lemma}

\begin{proof}
      For \(t=1\), the \(K_{t,t}\)-density hypothesis is the bound \(2|E(F)|\leq bk\) which is equivalent to \(d\le b\).
      Thus, we have \(d+1\leq b+1=s\) and for any weights \(0\le z_v \le p\) we have the bound \[\sum_vd_F(v)z_v\leq pbk\leq spk.
      \]
      This completes the proof for \(t=1\).

      Suppose \(t\geq2\), and write \(a=1-1/t\).
      For a set \(A\subseteq V(F)\) of size \(r\), write \(d_A(v)=|N_F(v)\cap A|\) and let \(c(S)\) be the number of common neighbors of a \(t\)-set \(S\).
      If \(r<t\), then \(|A|\le t-1\) and hence \[\sum_{v\in A}d_F(v)=\sum_vd_A(v)\leq(t-1)k.
      \]
      Otherwise, with \(R=\binom rt\), a standard double counting argument gives
      \begin{align}
            \sum_v\binom{d_A(v)}t                         & =\sum_{S\in\binom At}c(S), &
            \sum_{S\in\binom At}\binom{c(S)}t \leq2N_t(F) & \le bk/t.
      \end{align}
      Write \[U = \sum_v\binom{d_A(v)}t \] for the left-hand side of the first equation.
      In the chain of inequalities on the right, the factor \(2\) arises in the first inequality because each copy of \(K_{t,t}\) is counted at most twice, once for each part contained in \(A\).
      For every integer \(j\geq0\), note that \[ j\leq t-1+\left(t!
            \binom jt\right)^{1/t}
      \]
      because the term \(t-1\) suffices for all \(0\le j\le t-1\) and for \(j\ge t\) each factor in the product given by \(t! \binom{j}{t}\) is at least \(j-t+1\).
      For nonnegative \(y_1,\ldots,y_N\), H\"older's inequality gives \[ \sum_{i=1}^N y_i^{1/t} \leq N^{1-1/t}\left(\sum_{i=1}^N y_i\right)^{1/t}.
      \]
      First apply the integer inequality to each \(c(S)\), and then apply H\"older's inequality to the \(R=\binom{r}{t}\) terms indexed by \(S\in\binom At\), with \(y_S=\binom{c(S)}t\). This gives
      \begin{align*}
            U=\sum_{S\in\binom At}c(S) & \leq(t-1)R+(t!
                                                        )^{1/t}
            \sum_{S\in\binom At}\binom{c(S)}t^{1/t}                  \\
                                       & \leq(t-1)R+(t!)^{1/t}
            R^a \left(\sum_{S\in\binom At}\binom{c(S)}t\right)^{1/t} \\ & \leq(t-1)R+(t!
                                                                                         )^{1/t}
            R^a(bk/t)^{1/t}.
      \end{align*}
      Next, apply the integer inequality to each \(d_A(v)\), and apply H\"older to the \(k\) terms indexed by \(v\in V(F)\), with \(y_v=\binom{d_A(v)}t\) to obtain
      \begin{align*}
            \sum_{v\in A}d_F(v)
             & =\sum_{v\in V(F)}d_A(v)                         \\
             & \leq(t-1)k+(t!)^{1/t}
            \sum_{v\in V(F)}\binom{d_A(v)}t^{1/t}              \\
             & \leq(t-1)k+(t!)^{1/t}k^a
            \left(\sum_{v\in V(F)}\binom{d_A(v)}t\right)^{1/t} \\
             & =(t-1)k+(t!)^{1/t}k^aU^{1/t}.
      \end{align*}
      Using \(R = \binom rt\leq r^t/t!
      \) and the subadditivity of \(x^{1/t}\) gives
      \begin{align}
            \sum_{v\in A}d_F(v)
             & \le(t-1)k+(t-1)^{1/t}k^ar
            +\left(\frac{(t!)^2}{t}\right)^{1/t^2} b^{1/t^2}k^{a+1/t^2}r^a,
            \\&\le(t-1)k+(t-1)^{1/t}k^ar
            +t b^{1/t^2}k^{a+1/t^2}r^a,
      \end{align}
      since \(\bigl((t!)^2/t\bigr)^{1/t^2}\leq t\) for all positive integers \(t\).

      To bound the final term, use \(b\leq s^{2t-1}\) and \(a=1-1/t\) to obtain
      \begin{align*}
            t b^{1/t^2}k^{a+1/t^2}r^a
             & \leq t s^{(2t-1)/t^2}k^{a+1/t^2}r^a   \\
             & =t(sk)^{1/t}\bigl(s^{1/t}k^ar\bigr)^a \\
             & \leq sk+(t-1)s^{1/t}k^ar.
      \end{align*}
      The last step follows from a weighted AM-GM inequality (equivalently, Jensen's inequality for the logarithm).
      We start from the fact that for all \(\theta\in(0,1)\) and \(x,y\ge 0\) we have \(x^\theta y^{1-\theta}\le \theta x+(1-\theta)y\), and apply this with \(\theta=1/t\), \(x=sk\) and \(y=s^{1/t}k^ar\).
      Substituting in and using \(s\geq1\), and \((t-1)^{1/t}\leq t-1\), we conclude that
      \begin{equation}\label{eq:ktt-cut-budget}
            \sum_{v\in A}d_F(v)
            \leq tsk+2(t-1)s^{1/t}k^{1-1/t}|A|.
      \end{equation}
      This also holds when \(|A|<t\).
      To establish~\eqref{eq:ktt-weighted}, note that both sides of the desired inequality are linear in the weights \(z_v\), and hence it suffices to consider \(z_v\in\{0,p\}\).
      Taking \(A=\{v:z_v=p\}\), the desired fact~\eqref{eq:ktt-weighted} follows by multiplying the subset bound~\eqref{eq:ktt-cut-budget} above by \(p\).
      Taking \(A=V(F)\) in \eqref{eq:ktt-cut-budget} gives \[ d+1\leq 1+ts+2(t-1)s(k/s)^{1-1/t} \leq(3t-1)s\max\{1,(k/s)^{1-1/t}\}, \] which implies \eqref{eq:ktt-degree}.
\end{proof}

We now give explicit local occupancy parameters for graphs with hereditary \(K_{t,t}\)-density at most \(b\) in neighborhoods.
We state a finite sufficient condition and then give asymptotics that allow both \(b\) and \(t\) to grow.
In the parameter choice, we make use of the Lambert \(W\)-function, which is the nonnegative solution to \(W(x)e^{W(x)}=x\) for \(x\geq0\).

\begin{proposition}\label{prop:ktt-occupancy}
      Let \(\Delta\geq1\), let \(t\geq1\) be an integer, and let \(b\geq0\).
      Suppose that \(G\) has maximum degree at most \(\Delta\) and hereditary \(K_{t,t}\)-density at most \(b\) in neighborhoods.
      Set \(s=(b+1)^{1/(2t-1)}\), \(\rho=\Delta/s\), and \(D=\log\rho\).
      Suppose \(D\geq8et\), and define
      \begin{align}\label{eq:ktt-density-parameters}
            \lambda & =\frac1{sD^{2t-1}},               & M & =4esD^{2t}, &
            c       & =\exp\left(-\frac{4t(4e)^{1-1/t}}
            D\right).
      \end{align}
      Write \(p=\lambda/(1+\lambda)\).
      For any \(0\le z\le D\), \(G\) has local occupancy with parameters \((\lambda,\beta_z,\gamma_z)\) given by
      \begin{align}\label{eq:ktt-tangent-parameters}
            \beta_z  & =\frac{e^z}{p(1+z)}, &
            \gamma_z & =\frac1{c(1+z)}.
      \end{align}
      The choice \(z=W(p\Delta/c)\) satisfies \(0<z<D\).
      If \(D\to\infty\) and \(t\log D=o(D)\), then for this choice
      \begin{equation}\label{eq:ktt-Gamma}
            \Gamma=\beta_z+\gamma_z\Delta=\frac{\Delta}{cz}
            =\left(1+O\!\left(\frac{t\log D}{D}\right)\right)\frac{\Delta}{D}.
      \end{equation}
      The implicit constant is absolute, and the estimate is uniform in \(b\) and \(t\) throughout this regime.
\end{proposition}

\begin{proof}
      For the empty graph the local occupancy inequality is \(\beta_zp=e^z/(1+z)\geq1\).
      Let \(F\) be an induced subgraph of a neighborhood with \(k\geq1\) vertices and average degree \(d\).
      For any fugacity \(x>0\), we have the elementary bounds
      \begin{align}\label{eq:ktt-elementary}
            \log Z_F(x) & \leq k\log(1+x)\le kx,                  &
            E_F(x)      & \geq kx(1+x)^{-d-1}\geq kx e^{-(d+1)x}.
      \end{align}
      The first holds because independent sets are subsets of the vertex set, and the second is a simple conditioning and convexity argument that has been noted before, e.g.\ \cite[Lem.~6]{DdKP21} and~\cite[Lem.~21]{DKPS20a}.
      The identity \(xE_F'(x)=\operatorname{Var}_{\mu_{F,x}}(|I|)\geq0\) shows that \(E_F(x)\) is nondecreasing.
      Applying \eqref{eq:ktt-elementary} at
      \(y=\min\{\lambda,(d+1)^{-1}\}\) therefore gives
      \begin{equation}\label{eq:ktt-coarse}
            E_F(\lambda)\geq E_F(y) \geq\frac1e\min\{\lambda k,k/(d+1)\}.
      \end{equation}

      If \(k\leq M\), then \(M\geq s\) and \eqref{eq:ktt-degree} give \[ \lambda(d+1)\leq4t\lambda s(M/s)^{1-1/t} =\frac{4t(4e)^{1-1/t}}{D}.
      \]
      Consequently,
      \begin{equation}\label{eq:ktt-small}
            E_F(\lambda)\geq c\lambda k\geq c\log Z_F(\lambda).
      \end{equation}
      Write \(r=\log Z_F(\lambda)\) and for \(0\leq z\leq D\) fix \(\beta_z\), \(\gamma_z\) as above.
      Then \[ \beta_z\frac{p}{Z_F(\lambda)}+\gamma_zE_F(\lambda) \geq\frac{e^{z-r}+r}{1+z}\geq1, \] where the final inequality uses only \( e^{z-r} \ge 1+z-r\).

      If \(k>M\), then \eqref{eq:ktt-degree} and \eqref{eq:ktt-coarse} give \[ E_F(\lambda)\geq\frac1e\min\left\{ 4eD,\frac{(4e)^{1/t} D^2}{4t}\right\}\geq2D, \] where the last inequality follows from \(D\geq8et\).
      Since \(c(1+z)\leq1+D\leq2D\), the term involving \(E_F(\lambda)\) alone is at least one.

      Now take \(z=W(p\Delta/c)\).
      Since \(4t(4e)^{1-1/t}\leq16et\) and \(D\geq8et\), we have \(c^{-1}\leq e^2\), and hence \[ 0<\frac{p\Delta}{c}\leq\frac{e^{D+2}}{D^{2t-1}}<De^D.
      \]
      The monotonicity of \(W\) gives \(0<z<D\), and the identity \(ze^z=p\Delta/c\) yields \(\Gamma=\Delta/(cz)\).

      To obtain the asymptotic behavior, which is uniform in \(b\) and \(t\), suppose that \(D\to\infty\) and \(t\log D=o(D)\).
      Since \(s\geq1\), we have \(\lambda\leq D^{-(2t-1)}\leq D^{-1}\), so
      \begin{align*}
            \log(p\Delta/c)
             & =D-(2t-1)\log D-\log(1+\lambda)
            +\frac{4t(4e)^{1-1/t}}{D}          \\
             & =D-(2t-1)\log D+O(t/D).
      \end{align*}
      Set \(z_*=D-2t\log D\), which is eventually at least \(D/2\) under our assumptions.
      Then
      \begin{align*}
            z_*+\log z_*
             & =D-(2t-1)\log D+\log\left(1-\frac{2t\log D}{D}\right) \\
             & =D-(2t-1)\log D+O\!\left(\frac{t\log D}{D}\right).
      \end{align*}
      Comparing with \(z+\log z=\log(p\Delta/c)\), and using that the derivative of \(x+\log x\) is at least one for \(x>0\), gives \[ z=D-2t\log D+O\!
            \left(\frac{t\log D}{D}\right).
      \]
      All implicit constants above are absolute; the only bound used on \(s\) is \(s\geq1\), so the estimates are uniform in \(b\).
      Together with \(c^{-1}=1+O(t/D)\), this proves \eqref{eq:ktt-Gamma}.
\end{proof}

For fixed \(t\), the conclusion~\eqref{eq:ktt-Gamma} of \cref{prop:ktt-occupancy} gives \cref{thm:ktt-occupancy}, uniformly in \(b\).
It also gives the local occupancy required by the growing-\(t\) regime discussed in \cref{rem:ktt-growing-t}, since \(t\log D=o(D)\) implies that eventually \(D\geq8et\).

\section{The neighborhood block algorithm}\label{sec:block}

We analyze the neighborhood block kernel \(K^N\) defined in~\eqref{eq:dynamics-kernels}.
Samples can be obtained in time \(2^{O(\Delta)}\) per block step by enumerating the independent sets of the neighborhood.

We prove \cref{thm:block} without assuming that \(K^N\) is reversible or that it mixes rapidly (indeed, it may be that neither hold).
The following elementary entropy calculation is the discrete heat-bath version of the local stationarity argument in \cite{LMR+25}.
In particular, it is highly reminiscent of~\cite[Lem.~3.2]{LMR+25}.

\begin{lemma}\label{lem:entropy}
      For every vertex \(u\) and every probability distribution \(\nu\) on \(\cI(G)\),
      \begin{equation}\label{eq:entropy-identity}
            \KL(\nu\Vert\pi)-\KL(\nu P_u\Vert\pi)
            =\KL(\nu\Vert\nu P_u)
            \geq2\tv(\nu,\nu P_u)^2.
      \end{equation}
      The same identity and inequality hold with \(P_u\) replaced by \(Q_u\).
\end{lemma}

\begin{proof}
      Apply the chain rule for relative entropy, conditioning on the spins in \(V\setminus\{u\}\) for \(P_u\), or in \(V\setminus N(u)\) for \(Q_u\).
      In either case, the initial and updated probability distributions have the same exterior marginal.
      That is, for \(P_u\) note that both \(\nu\) and \(\nu P_u\) yield the same distribution on \(V\setminus\{u\}\), and for \(Q_u\) both \(\nu\) and \(\nu Q_u\) yield the same distribution on \(V\setminus N(u)\).
      Moreover, the heat-bath kernels resample \(u\) or \(N(u)\) according to the same conditional distribution as \(\pi\) given the exterior, so the conditional relative entropy is zero.
      Subtracting the two chain-rule expressions gives the equality.
      Pinsker's inequality gives the last bound.
\end{proof}

\begin{proof}[Proof of \cref{thm:block}]
      For an arbitrary probability distribution \(\nu\) on the independent sets of \(G\), write
      \begin{align}
            a_u          & =\tv(\nu,\nu Q_u),                      &
            b_u          & =\tv(\nu Q_u,\nu Q_uP_u),               &
            D_{K^N}(\nu) & =\KL(\nu\Vert\pi)-\KL(\nu K^N\Vert\pi).
      \end{align}
      Convexity of relative entropy and two applications of
      \cref{lem:entropy} show that
      \begin{equation}\label{eq:block-dissipation}
            D_{K^N}(\nu)\geq\frac2n\sum_u(a_u^2+b_u^2).
      \end{equation}
      Let \(X_u(I) = \indicator{u\in I}\) and let \(Y_u(I) = |N(u)\cap I|\) be functions on \(\cI(G)\) that we interpret as random variables when we have a measure on \(\cI(G)\).
      The neighborhood update \(Q_u\) leaves \(X_u\) unchanged, while the center update \(P_u\) leaves \(Y_u\) unchanged.
      Since these functions take values in \([0,1]\) and \([0,d_G(u)]\), respectively, local occupancy in the form~\eqref{eq:block-certificate} gives \[ 1\leq\beta\E_{\nu Q_uP_u}X_u+\gamma\E_{\nu Q_u}Y_u \leq\beta\E_\nu X_u+\gamma\E_\nu Y_u+\gamma d_G(u)a_u+\beta b_u.
      \]
      We sum over \(u\) and apply the Cauchy--Schwarz inequality and \(d_G(u)\le \Delta\) to obtain
      \begin{equation}\label{eq:block-local-stationarity}
            1\leq\frac{\Gamma}{n}\E_\nu|I|
            +\sqrt{\frac{(\beta^2 + \gamma^2\Delta^2)D_{K^N}(\nu)}2}.
      \end{equation}
      Let \(\nu_t=\nu_0(K^N)^t\).
      By a telescoping sum, for any time \(T\ge 1\), \[\sum_{t=0}^{T-1}D_{K^N}(\nu_t)= \KL(\nu_0\Vert\pi) - \KL(\nu_T\Vert\pi) \leq \KL(\nu_0\Vert\pi).
      \]
      Since \(J\) is uniform on \(\{0,\ldots,T-1\}\) and independent of the chain,
      \[
            \E|I_J|=\frac1T\sum_{t=0}^{T-1}\E_{\nu_t}|I|.
      \]
      Averaging \eqref{eq:block-local-stationarity} over these times and applying Cauchy--Schwarz again gives
      \begin{align*}
            1
             & \leq\frac{\Gamma}{n}\E|I_J|
            +\sqrt{\frac{(\beta^2 + \gamma^2\Delta^2)}{2}}\,
            \frac1T\sum_{t=0}^{T-1}\sqrt{D_{K^N}(\nu_t)}  \\
             & \leq\frac{\Gamma}{n}\E|I_J|
            +\sqrt{\frac{(\beta^2 + \gamma^2\Delta^2)}{2T}
                   \sum_{t=0}^{T-1}
                   D_{K^N}(\nu_t)} \\ &\leq\frac{\Gamma}{n}\E|I_J| +\sqrt{\frac{(\beta^2 + \gamma^2\Delta^2)H_0}{2T}}.
      \end{align*}
      The last step uses the entropy bound above and \(H_0=\KL(\nu_0\Vert\pi)\).
      Rearranging gives~\eqref{eq:block-bound}.
\end{proof}

\begin{remark}
      \label{rem:amplification}
      For certificate error \(0\leq\eta<1\), choose \(0<\epsilon<1-\eta\) and \(0<\delta<1\), arrange an expected-size guarantee \((1-\eta-\epsilon/2)n/\Gamma\), and retain the largest set visited in the run.
      We can take \(\eta=0\) for \cref{thm:block} and our applications.
      If \(Y\) is the largest size seen, then \(0\leq Y\leq n\), and \[ \Pr\left(Y\geq(1-\eta-\epsilon)n/\Gamma\right)\geq\frac{\epsilon}{2\Gamma}.
      \]
      Indeed, \(\E Y\leq(1-\eta-\epsilon)n/\Gamma+n\Pr(Y\geq(1-\eta-\epsilon)n/\Gamma)\).
      Taking the largest output of \(\lceil(2\Gamma/\epsilon)\log(1/\delta)\rceil\) independent runs therefore gives size at least \((1-\eta-\epsilon)n/\Gamma\) with probability at least \(1-\delta\).
      The same elementary amplification argument applies to all our dynamics guarantees.
\end{remark}

\section{Transfer to ordinary Glauber dynamics}\label{sec:transfer}

We now consider the ordinary Glauber dynamics kernel \(K^G\) from \eqref{eq:dynamics-kernels}.
Use the words \(S_u\) and the occurrence bound \(\kappa\) from \cref{thm:transfer}, and write \[ M_{uv}=|\{i:v_{u,i}=v\}|, \] so that \(\max_{v\in V}\sum_u M_{uv}=\kappa\).
If every word has length at most \(L\), then \(\kappa\leq nL\).
For \(k\) full sweeps of each neighborhood, \(M_{uv}=k\indicator{v\in N(u)}\), so \(\kappa\leq k\Delta\).

\begin{proof}[Proof of \cref{thm:transfer}]
      For a probability distribution \(\nu\) on the independent sets of \(G\), for \(u\in V\) set \(a_u=\tv(\nu,\nu P_u)\).
      By the triangle inequality, total variation distance can only decrease under Markov kernels.
      That is, for any distributions \(\mu, \mu'\) we have \(\tv(\mu K^G,\mu' K^G)\leq\tv(\mu,\mu')\).
      This gives control over a telescoping product of kernels in the sense that
      \[ \nu S_u-\nu=\sum_{i=1}^{\ell_u} (\nu P_{v_{u,i}}-\nu)(P_{v_{u,i+1}}\cdots P_{v_{u,\ell_u}}),
      \]
      and we apply TV-contraction at every term of the sum to obtain
      \begin{equation}\label{eq:word-telescoping1}
            \tv(\nu,\nu S_u)\leq \sum_{i=1}^{\ell_u} \tv(\nu P_{v_{u,i}},\nu) \leq\sum_v M_{uv}a_v.
      \end{equation}
      To compare occupancy scores, define \[ f_u(I)=\beta p\indicator{I\cap N(u)=\emptyset}+\gamma|I\cap N(u)|, \qquad L_u=\max\{\beta p,\gamma d_G(u)\}.
      \]
      The two summands in \(f_u\) cannot both be positive, so \(0\leq f_u\leq L_u\).
      Standard conditional probability in the hard-core model gives \(\Pr_{\nu S_uP_u}(u\in I)=p\Pr_{\nu S_u}(I\cap N(u)=\emptyset)\), so the certificate \eqref{eq:word-certificate} says that \(\E_{\nu S_u}f_u\geq1-\eta\).
      For a function \(f\) taking values in \([0,L]\), we have \( |\E_\mu f-\E_\nu f|\leq L\tv(\mu,\nu) \).
      Apply this first to \(f_u\) using \eqref{eq:word-telescoping1}, and then to the indicator of \(u\in I\) using \(a_u=\tv(\nu,\nu P_u)\), to obtain
      \begin{align*}
            1-\eta & \leq\E_{\nu S_u}f_u
            \leq\E_\nu f_u+L_u\sum_v M_{uv}a_v                                                        \\
                   & =\beta\Pr_{\nu P_u}(u\in I)+\gamma\E_\nu|I\cap N(u)|+L_u\sum_v M_{uv}a_v         \\
                   & \leq\beta\Pr_\nu(u\in I)+\gamma\E_\nu|I\cap N(u)|+\beta a_u+L_u\sum_v M_{uv}a_v.
      \end{align*}
      We recall the familiar identity and bounds
      \begin{align*}
            \sum_u\Pr_\nu(u\in I) & =\E_\nu|I|,                                    \\
            \sum_u\E_\nu|I\cap N(u)|
                                  & =\E_\nu\sum_{v\in I}d_G(v)\leq\Delta\E_\nu|I|.
      \end{align*}
      Averaging over a uniform random \(u\) and collecting the coefficient of each \(a_v\), we obtain (with \(\Gamma = \beta+\Delta\gamma\))
      \begin{align}
            1-\eta
             & \leq\frac{\Gamma}{n}\E_\nu|I|
            +\frac1n\sum_v\left[\beta+
                                \sum_u L_uM_{uv}\right]a_v\notag \\
             & \leq\frac{\Gamma}{n}\E_\nu|I|+C\frac1n\sum_v a_v.
            \label{eq:word-occupancy-error}
      \end{align}
      Here, \(L_u\leq\max\{\beta p,\gamma\Delta\}\) and \(\sum_u M_{uv}\leq\kappa\), so \[ \beta+\sum_u L_uM_{uv} \leq\beta+\max\{\beta p,\gamma\Delta\}\sum_u M_{uv} \leq\beta+\kappa\max\{\beta p,\gamma\Delta\}=C.
      \]

      For the alternative setting where we instead assume an average local occupancy certificate~\eqref{eq:averaged-word-certificate}, each prefix has at most \(M_{uv}\) occurrences of \(v\).
      Apply the same bounds to each prefix and average; this gives \eqref{eq:word-occupancy-error} unchanged.

      On the other hand, convexity and \cref{lem:entropy} imply \[ D_{K^G}(\nu):=\KL(\nu\Vert\pi)-\KL(\nu K^G\Vert\pi) \geq\frac2n\sum_v a_v^2.
      \]
      By the Cauchy--Schwarz inequality we therefore have \(n^{-1}\sum_v a_v\leq\sqrt{D_{K^G}(\nu)/2}\).
      For \(\nu_t=\nu_0(K^G)^t\), sum \eqref{eq:word-occupancy-error} over \(0\leq t<T\) and use \(\sum_t D_{K^G}(\nu_t)\leq H_0\).
      This gives \eqref{eq:glauber-bound}.
      Since \(H_0\leq H_\lambda\), taking \(T\geq C^2H_\lambda/(2\epsilon^2)\) makes the error term at most \(\epsilon\), giving expected size at least \((1-\eta-\epsilon)n/\Gamma\) as required.
\end{proof}

The following remark shows that local occupancy plus a short sequence of single-site updates which corresponds to exact or approximate (in the sense of total variation) conditional resampling of the neighborhood implies our condition.
This is intuitively clear, since the equilibrium measure satisfies local occupancy and hence a short sequence of updates which implements exact or approximate conditional sampling achieves the required local occupancy guarantee in expectation.

\begin{remark}
      Suppose that \(G\) satisfies local occupancy with parameters \((\lambda,\beta,\gamma)\) and, for some \(\theta\geq0\), the words \(S_u\) satisfy
      \begin{equation}\label{eq:tv-simulation}
            \tv\bigl(S_u(I,\cdot),Q_u(I,\cdot)\bigr)\leq\theta
      \end{equation}
      for every independent set \(I\) and every vertex \(u\).
      Then \eqref{eq:word-certificate} holds with \(\eta=\max\{\beta p,\gamma\Delta\}\theta\).
      Whenever \(\eta<1\), \cref{thm:transfer} therefore gives the corresponding guarantee for ordinary Glauber dynamics, and if the sequence \(S_u\) corresponds exactly to the conditional sampling kernel \(Q_u\) we have \(\theta=0\) and thus \(\eta=0\).

      Indeed, convexity of total variation implies that \eqref{eq:tv-simulation} also gives \(\tv(\nu S_u,\nu Q_u)\leq\theta\) for every input probability distribution \(\nu\).
      For the score function \(f_u\) in the proof above, \eqref{eq:block-certificate} gives \(\E_{\nu Q_u}f_u\geq1\).
      Since \(0\leq f_u\leq\max\{\beta p,\gamma d_G(u)\}\), replacing \(Q_u\) by \(S_u\) gives \[ \beta\Pr_{\nu S_uP_u}(u\in I)+\gamma\E_{\nu S_u}|I\cap N(u)| =\E_{\nu S_u}f_u \geq1-\max\{\beta p,\gamma d_G(u)\}\theta\geq1-\eta.
      \]
\end{remark}

\section{Establishing neighborhood sweep conditions}\label{sec:sweeps}

Recall that we define the maximum average degree of a graph \(H\) by \[ \operatorname{mad}(H)= \max_{\emptyset\ne A\subseteq V(H)}\frac{2|E(H[A])|}{|A|}, \qquad \operatorname{mad}(\emptyset)=0.
\]
We say that \(G\) has local maximum average degree at most \(a\) if
\(\operatorname{mad}(G[N(u)])\leq a\) for every \(u\).
This is equivalent to the case \(t=1\) and \(b=a\) of the hereditary \(K_{t,t}\)-density condition in \cref{prop:ktt-occupancy}.

For \(\Delta\geq1\) and \(a\geq0\), introduce
\begin{align}\label{eq:mad-parameters}
      \ell   & =\log(1+\lambda),                  &
      D      & =\Delta(1+\lambda)^a\ell,          &
      w      & =W(D),                             &
      \beta  & =\frac{e^w}{p(1+w)},               &
      \gamma & =\frac{(1+\lambda)^a\ell}{p(1+w)}.
\end{align}
Here \(W\) is the Lambert function introduced in \cref{prop:ktt-occupancy}.
In particular,
\begin{equation}\label{eq:mad-Gamma}
      \Gamma=\beta+\gamma\Delta=\frac{e^w}{p}=\frac{\Delta(1+\lambda)^a\log(1+\lambda)}{pW\bigl(\Delta(1+\lambda)^a\log(1+\lambda)\bigr)}
      .
\end{equation}
These are the local occupancy parameters introduced in~\cite{DdKP21} and used again in~\cite{DKPS20a,DKPS20b}.

The following theorem gives an alternative to the triangle-free analysis of~\cite{LMR+24,LMR+25} and extends it to graphs with bounded local mad via local occupancy parameters from~\cite{DdKP21}.

\begin{theorem}
      \label{thm:sweeps}
      The following statements hold for a graph \(G\) with maximum degree at most \(\Delta\) and any fugacity \(\lambda>0\).
      \begin{enumerata}
            \item If \(G\) is triangle-free, one sweep of \(N(u)\) in any fixed
            order exactly implements \(Q_u\), for every exterior configuration
            and every initial state.
            Thus, the local occupancy guarantee is obtained with ordinary Glauber dynamics by \cref{thm:transfer} with \(\eta=0\) and \(C=\beta+\Delta\max\{\beta p,\gamma\Delta\}\).
            \item If \(G\) has local maximum average degree at most \(a\), it satisfies local occupancy with the
            parameters \((\lambda,\beta,\gamma)\) given in \eqref{eq:mad-parameters}.
            Two full sweeps of \(N(u)\), each in a fixed order, satisfy \eqref{eq:word-certificate} with these same parameters and \(\eta=0\).
            Consequently, with time bound \[ T\geq\frac{(\beta+2\Delta\max\{\beta p,\gamma\Delta\})^2H_\lambda}{2\epsilon^2} \] and \(J\) uniform on \(\{0,\ldots,T-1\}\), Glauber dynamics from any initial state satisfies \[ \E|I_J|\geq(1-\epsilon)n/\Gamma.
            \]
            This is the same bound as the corresponding equilibrium
            local-occupancy estimate, up to the error factor \(1-\epsilon\).
      \end{enumerata}
\end{theorem}

The fact that a single sweep implements exact conditional resampling in neighborhoods of triangle-free graphs is trivial since the hard-core measure on a graph with no edges is a product measure.
To prove the rest of the theorem we need the following lemma, which establishes the occupancy estimates after two sweeps of a neighborhood with bounded average degree.
This result serves as a warmup to the more complex analysis of the hereditary \(K_{t,t}\)-density condition in \cref{sec:ktt-sweeps}.
The proof is essentially an interpretation of the proofs of~\cite[Lem.~6]{DdKP21} and~\cite[Lem.~21]{DKPS20a} in the context of heat-bath dynamics.

\begin{lemma}
      \label{lem:two-sweeps}
      Let \(F\) have \(k\) vertices and average degree at most \(a\).
      Starting from any independent set, apply two full heat-bath sweeps, with deterministic orders, at fugacity \(\lambda\).
      For the resulting independent set \(Y\),
      \begin{equation}\label{eq:sweep-moments}
            \Pr(Y=\emptyset)\geq(1+\lambda)^{-k},\qquad
            \E|Y|\geq p\sum_{v\in V(F)}(1+\lambda)^{-d_F(v)}
            \geq pk(1+\lambda)^{-a}.
      \end{equation}
\end{lemma}

\begin{proof}
      The bounds are immediate if \(k=0\), so suppose \(k>0\).
      A vertex has spin \(1\) when it is occupied and spin \(0\) when it is unoccupied.
      For each vertex \(v\), let \(B_{v,1}\) and \(B_{v,2}\) be the proposal coins for its updates in the first and second sweeps, respectively.
      All these coins are independent Bernoulli-\(p\) random variables, where \(p=\lambda/(1+\lambda)\).
      At the update of \(v\) in sweep \(i\), we set its spin to \(B_{v,i}\) if all its neighbors are unoccupied, and to \(0\) otherwise.
      This implements the heat-bath update.
      In particular, a zero proposal always leaves the vertex unoccupied, and its spin stays unchanged until its next update.

      If \(B_{v,2}=0\) for every vertex \(v\), every vertex is unoccupied at the end of the second sweep.
      Thus \[ \Pr(Y=\emptyset)\geq(1-p)^k=(1+\lambda)^{-k}.
      \]

      To bound the expected size of \(Y\), fix a vertex \(v\) and consider the moment just before its update in the second sweep.
      A neighbor \(w\) that comes before \(v\) in this sweep has already received its second update, so it is unoccupied if \(B_{w,2}=0\).
      A neighbor \(w\) that comes after \(v\) still has its spin from the end of the first sweep, so it is unoccupied if \(B_{w,1}=0\).
      This is the role of the first sweep: even neighbors still awaiting their second update have a spin controlled by a proposal coin, regardless of the initial independent set.

      We have therefore identified one coin for each neighbor of \(v\) such that the coin being zero guarantees that this neighbor is unoccupied when \(v\) is updated.
      The choice of coins depends only on the fixed sweep orders, so these \(d_F(v)\) coins are independent of one another and of \(B_{v,2}\).
      If they are all zero and \(B_{v,2}=1\), then \(v\) becomes occupied.
      Since \(v\) is not updated again, it belongs to \(Y\).
      Consequently, \[ \Pr(v\in Y)\geq p(1-p)^{d_F(v)} =p(1+\lambda)^{-d_F(v)}.
      \]
      Summing over the vertices gives the first lower bound on \(\E|Y|\).
      For the second, write \(\bar d=k^{-1}\sum_v d_F(v)\leq a\) for the average degree.
      Since \(t\mapsto(1+\lambda)^{-t}\) is convex and decreasing, Jensen's inequality gives \[ \frac1k\sum_{v\in V(F)}(1+\lambda)^{-d_F(v)} \geq(1+\lambda)^{-\bar d} \geq(1+\lambda)^{-a}, \] as required.
\end{proof}

\begin{proof}[Proof of \cref{thm:sweeps}]
      Fix a vertex \(u\) and an exterior spin configuration \(J\) on
      \(V\setminus N(u)\).
      If \(G\) is triangle-free, the available graph \(F(u,J)\) is edgeless.
      Each available vertex receives an independent Bernoulli-\(p\) spin when updated; all unavailable vertices stay zero.
      One sweep therefore produces exactly the conditional hard-core measure, independently of its initial state.
      There is one occurrence of \(v\) for every center adjacent to \(v\), so \(\kappa\leq\Delta\).
      This proves part (a).

      For part (b), let \(F=F(u,J)\) and \(k=|V(F)|\).
      Its average degree is at most \(a\).
      Under the fixed exterior, unavailable vertices always get spin zero, so the two neighborhood sweeps restrict to two full sweeps on \(F\).
      After them, write \(Y\) for the occupied part of the neighborhood.
      Updating \(u\) now gives occupation probability \(p\Pr(Y=\emptyset)\).
      \Cref{lem:two-sweeps} bounds the left side of the required short update sequence condition~\eqref{eq:word-certificate} from below by
      \begin{equation}\label{eq:mad-certificate}
            \beta p e^{-\ell k}+\gamma pk e^{-a\ell}
            =\frac{e^{w-\ell k}+\ell k}{1+w}\geq1.
      \end{equation}
      The last inequality is simply \(e^x\geq1+x\) with \(x=w-\ell k\).
      It also covers the empty available graph.
      Since both the configuration outside \(N(u)\) and the initial independent set were arbitrary, \eqref{eq:word-certificate} holds with \(\eta=0\).
      Finally, \(we^w=D\) gives \eqref{eq:mad-Gamma}, and we have \(\kappa\leq2\Delta\) in the case of words corresponding to two full sweeps of each neighborhood.
      Applying \cref{thm:transfer} completes the proof.
\end{proof}

It is not too hard to deduce the consequences of our analysis in the case of a maximum average degree bound \(a\) which grows with the maximum degree \(\Delta\).

\begin{corollary}
      \label{cor:asymptotic}
      Let \(a=a(\Delta)\geq0\) and write \(R=\Delta/(a+1)\).
      For graphs of maximum degree at most \(\Delta\) and local maximum average degree at most \(a\), taking \(\lambda=1/((a+1)\log R)\) gives, uniformly in \(a\) as \(R\to\infty\), for the parameters~\eqref{eq:mad-Gamma} \[ \frac{1}{\Gamma} =\left(1-O\!
            \left(\frac{\log\log R}{\log R}\right)\right)
            \frac{\log R}{\Delta}.
      \]
      In particular, \(a+1=\Delta^{o(1)}\) gives size \((1-o(1))n\log\Delta/\Delta\).
\end{corollary}

\begin{proof}
      Uniformly in \(a\), we have \(\lambda\leq1/\log R\) and \(a\lambda\leq1/\log R\), and hence the parameters introduced in \eqref{eq:mad-parameters} for this setting satisfy
      \begin{align}
            \frac{p}{\ell} & =1+O(1/\log R),                             &
            (1+\lambda)^a  & =1+O(1/\log R),                             &
            D              & =\bigl(1+O(1/\log R)\bigr)\frac{R}{\log R}.
      \end{align}
      Since \(W(D)+\log W(D)=\log D\), we have \(W(D)=\log R-O(\log\log R)\).
      The result now follows from~\eqref{eq:mad-Gamma}.
\end{proof}

For triangle-free graphs this conclusion is consistent with the small-fugacity strengthening described in \cite[Rem.~1.5]{LMR+25}.
The two-sweep argument explains why the same approach works for growing local maximum average degree: the last proposal coins provide the required neighborhood estimates even though the spins themselves are dependent.

\section{Three sweeps and hereditary \texorpdfstring{\(K_{t,t}\)}{Ktt} density}
\label{sec:ktt-sweeps}

In this section we adopt the local occupancy parameters and notation of \cref{prop:ktt-occupancy}.
We recall \(s=(b+1)^{1/(2t-1)}\), \(\rho=\Delta/s\), and \(D=\log\rho\).
We work in the regime \(D\to\infty\) and \(t\log D=o(D)\), allowing both \(b\) and \(t\) to vary; this includes fixed \(t\) with \(\rho\to\infty\).
The parameters are
\begin{align}
      \lambda & =\frac1{sD^{2t-1}},               & M & =4esD^{2t}, &
      c       & =\exp\left(-\frac{4t(4e)^{1-1/t}}
      D\right).
\end{align}
With \(p=\lambda/(1+\lambda)\), we also set \(z=W(p\Delta/c)\) and
\begin{align}
      \beta  & =\frac{e^z}{p(1+z)}, &
      \gamma & =\frac1{c(1+z)}.
\end{align}
For these parameters, \(\Gamma=\beta+\gamma\Delta=\Delta/(cz)\), as in \eqref{eq:ktt-Gamma}.
For each \(u\), fix an arbitrary ordering of the neighborhood \(N(u)\) and write \(W_u\) for the kernel consisting of one full sweep of single-vertex heat-bath updates in that order.
For the analysis, we consider performing exactly three sweeps, corresponding to the kernel
\begin{equation}\label{eq:ktt-word}
      S_u=W_u^3.
\end{equation}
The exterior configuration \(J\) on \(V\setminus N(u)\) (including the spin of \(u\)) is fixed throughout.
At the end of the second and third sweeps, writing \(I\) for the current independent set, record the \emph{scores} \[ g_u(I)=\beta p\indicator{I\cap N(u)=\emptyset}+\gamma |N(u)\cap I|.
\]
The function \(g_u(I)\) is the expected local score after modifying \(I\) with a hypothetical heat-bath update at \(u\).
We analyze the arithmetic average of these two recorded scores:
\begin{equation}\label{eq:ktt-certificate}
      \frac12\left(\E_{\nu W_u^2}g_u+\E_{\nu W_u^3}g_u\right)\geq1.
\end{equation}
This is \eqref{eq:averaged-word-certificate} for two fixed prefixes of the three-sweep word corresponding to two and three full sweeps, respectively.

The analysis of the sweeps mirrors the local occupancy argument in \cref{prop:ktt-occupancy} in which we rely on different estimates of the score for small and large available graphs.
For small available graphs, the estimates after two sweeps already suffice.
For large ones, we replace the equilibrium monotonicity in fugacity used in \eqref{eq:ktt-coarse} by the weighted degree bound offered by \cref{lem:ktt-weighted}.

\begin{lemma}
      \label{lem:ktt-sweep-mean}
      Let \(t\geq1\) be an integer, let \(b\geq0\), and let \(F\) have \(k\geq1\) vertices and \(K_{t,t}\)-density at most \(b\).
      Set \(s=(b+1)^{1/(2t-1)}\).
      Starting from any independent set in \(F\), perform three full heat-bath sweeps at fugacity \(\lambda>0\), each in a fixed order, and let \(X\) and \(Y\) be the sets after the second and third sweeps.
      Write \(p=\lambda/(1+\lambda)\).
      Then
      \begin{equation}\label{eq:ktt-sweep-mean}
            \frac{\E|X|+\E|Y|}{2}
            \geq\frac{pk(1-8tsp)}{2(1+4tps^{1/t}k^{1-1/t})}.
      \end{equation}
\end{lemma}

\begin{proof}
      Write \(x_v=\Pr(v\in X)\) and \(y_v=\Pr(v\in Y)\).
      We implement each heat-bath update using a fresh, independent Bernoulli-\(p\) random variable as a proposal: the updated vertex is occupied exactly when the proposal is \(1\) and all its neighbors are unoccupied.
      At either recorded endpoint, the spin of each vertex is therefore at most its proposal in that sweep, so \(0\leq x_v,y_v\leq p\).

      For each vertex \(v\), let \(I_v^-\) be the occupied set just before its update in the third sweep.
      Vertices already updated in this sweep have their spins from \(Y\), while those awaiting their update still have their spins from \(X\).
      Hence \(I_v^-\subseteq X\cup Y\) for every realization of the updates, and a union bound gives \[ \Pr\bigl(N_F(v)\cap I_v^-\ne\emptyset\bigr) \leq\sum_{w\in N_F(v)}(x_w+y_w).
      \]
      The proposal at this update of \(v\) is independent of \(I_v^-\), and \(v\) is not updated again before the end of the sweep.
      Consequently, \[ y_v=p\Pr\bigl(N_F(v)\cap I_v^-=\emptyset\bigr) \geq p\left(1-\sum_{w\in N_F(v)}(x_w+y_w)\right).
      \]
      Note that these resemble the same calculations as for the two-sweep argument but we are now considering a third sweep in addition.
      Summing over \(v\), each term \(x_w+y_w\) is counted \(d_F(w)\) times, giving \[ \E|Y|= \sum_v y_v \geq pk-p\sum_{w\in V(F)}d_F(w)(x_w+y_w).
      \]

      Set \(H=\E|X|+\E|Y|\).
      Apply \cref{lem:ktt-weighted} separately with the vertex weights given by \((x_w)_{w\in V(F)}\) and \((y_w)_{w\in V(F)}\), which lie in \([0,p]\) as required, to obtain \[ \sum_{w\in V(F)}d_F(w)(x_w+y_w) \leq8tspk+4ts^{1/t}k^{1-1/t}H.
      \]
      Combining these bounds with \(\E|Y|\leq H\) gives
      \[
            \bigl(1+4tps^{1/t}k^{1-1/t}\bigr)H
            \geq pk(1-8tsp).
      \]
      Using \(H=\E|X|+\E|Y|\) and rearranging gives~\eqref{eq:ktt-sweep-mean}, as required.
\end{proof}

\begin{proof}[Proof of \cref{cor:ktt-glauber}]
      Fix a vertex \(u\) and an exterior spin configuration representing an independent set \(J\subset V(G)\setminus N(u)\).
      Since \(G\) has hereditary \(K_{t,t}\)-density at most \(b\) in neighborhoods, the available graph \(F=F(u,J)\) has \(K_{t,t}\)-density at most \(b\) and some number of vertices \(k\ge 0\).
      Unavailable vertices have their spin fixed to zero, so full neighborhood sweeps restrict to sweeps on the vertices of \(F\).
      Write \(Y_2\), \(Y_3\) for the sets of occupied vertices in \(F\) after the second and third sweeps respectively.
      The expected average score is \[ \frac{\beta p}{2}\bigl(\Pr(Y_2=\emptyset)+\Pr(Y_3=\emptyset)\bigr) +\frac{\gamma}{2}\bigl(\E|Y_2|+\E|Y_3|\bigr).
      \]

      For \(0\le k\le M\) we do not really need the third sweep, and the bounds we need follow from the choice of parameters and \cref{lem:two-sweeps}.
      For \(k=0\) the expected average score is exactly \(\beta p=e^z/(1+z)\geq1\).
      If \(1\leq k\leq M\), apply \cref{lem:two-sweeps} to the first two sweeps to obtain its bounds for \(Y_2\).
      For \(Y_3\), condition on the independent set after the first sweep and apply the lemma to the second and third sweeps.
      Since the bounds hold for every initial independent set, they also hold after averaging over this conditioning.
      With \(d\) the average degree of \(F\), \eqref{eq:ktt-degree} gives, for \(j\in\{2,3\}\), \[ \Pr(Y_j=\emptyset)\geq(1+\lambda)^{-k}\geq e^{-\lambda k}, \qquad \E|Y_j|\geq\lambda k(1+\lambda)^{-d-1}\geq c\lambda k.
      \]
      Consequently, the expected score at each endpoint is at least
      \[
            \beta p e^{-\lambda k}+\gamma c\lambda k
            =\frac{e^{z-\lambda k}+\lambda k}{1+z}\geq1.
      \]
      The average has the same lower bound.

      If \(k>M\), we use \cref{lem:ktt-sweep-mean}.
      Since \(8tsp\leq8t/D=o(1)\), we eventually have \(1-8tsp>0\), and \(x\mapsto px/(1+4tps^{1/t}x^{1-1/t})\) is increasing on \((0,\infty)\).
      Hence by \cref{lem:ktt-sweep-mean} we have \[ \frac{\E|Y_2|+\E|Y_3|}{2} \geq\frac{pM(1-8tsp)}{2(1+4tps^{1/t}M^{1-1/t})} =\left(2e+O\!
            \left(\frac tD\right)\right)D\geq2D.
      \]
      The error has an absolute implicit constant, uniformly in \(b\) and \(t\) in the stated regime: \(\lambda\leq D^{-1}\) and
      \[
            pM=\frac{4eD}{1+\lambda},\qquad sp\leq D^{-(2t-1)},\qquad
            ps^{1/t}
            M^{1-1/t}\leq\frac{(4e)^{1-1/t}}D\leq\frac{4e}D.
      \]
      By \cref{prop:ktt-occupancy}, \(0<z<D\) once \(D\geq8et\), which holds eventually in this regime.
      Since \(0<c\leq1\) and \(D\geq1\), we have \(c(1+z)\leq1+D\leq2D\).
      Using the lower bound on their expected number just proved and the definition \(\gamma=1/[c(1+z)]\), we obtain \[ \frac{\gamma}{2}\bigl(\E|Y_2|+\E|Y_3|\bigr) \geq2D\gamma =\frac{2D}{c(1+z)} \geq1.
      \]
      The remaining term, involving the probabilities of an empty neighborhood, is nonnegative, so the expected average score is at least one.
      This proves \eqref{eq:ktt-certificate} from every input, with \(\eta=0\).

      Each neighbor is updated exactly three times, giving \(\kappa\leq3\Delta\).
      The averaged-prefix form of \cref{thm:transfer} (i.e.\ under hypothesis~\eqref{eq:averaged-word-certificate}) now gives the bound on expected size in \cref{cor:ktt-glauber}.
      These estimates also handle the growing-\(t\) regime discussed in \cref{rem:ktt-growing-t}.
      Indeed, the factors \(8tsp\) and \(4tps^{1/t}M^{1-1/t}\) are both \(O(t/D)=o(1)\), so when \(k>M\) the average expected number of occupied vertices after the second and third sweeps is at least \((2e+o(1))D\).
      Both for fixed \(t\) and the discussed range of growing \(t\), we have \(\Gamma=O(\Delta/D)\) and \(H_\lambda=O(n(\log s+t\log D))\), with absolute implicit constants.
      A sufficient number of single-site updates is therefore \[ O\!
            \left(1+
            \frac{n\Delta^4(\log s+t\log D)}{\epsilon^2D^2}\right).
      \]
      For the high probability bound, take \(\epsilon=1/D\), use the asymptotic form~\eqref{eq:ktt-Gamma} of \(\Gamma\), and apply
      \cref{rem:amplification}.
      Since \(1\leq s\leq\Delta\leq n\) and \(\log s+t\log D=O(\log\Delta)\) in both regimes, the update and repetition counts are polynomially bounded, and each update costs \(O(\Delta)\) time.

      For implementation with random bits, put \(x=sD^{2t-1}\), choose an integer \(q\) with \(x\leq q\leq x+2\), and use the rational fugacity \(\widehat\lambda=1/q\).
      Such an integer can be found by approximating \(x\) within \(1/4\) and rounding the resulting upper bound upward.
      In both regimes, \((2t-1)\log D\leq D\) eventually, so \(x\leq se^D=\Delta\) and \(q\leq\Delta+2\).
      The proposal probability is then \(\widehat p=1/(q+1)\), so exact sampling by rejection uses an expected \(O(\log n)\) random bits per update.
      Recompute \(p,z,\beta_z,\gamma_z\) using \(\widehat\lambda\), while retaining \(c,M,D\).
      Uniformly in \(b\), we have \(\widehat\lambda/\lambda=1-O(D^{-1})\).
      For \(1\leq k\leq M\), the bounds on \(\Pr(Y_j=\emptyset)\) and \(\E|Y_j|\), \(j\in\{2,3\}\), remain valid with \(\lambda\) replaced by \(\widehat\lambda\), since \(\widehat\lambda(d+1)\leq\lambda(d+1)\) and the case \(k=0\) is immediate.
      For \(k>M\), the lower bound on \((\E|Y_2|+\E|Y_3|)/2\) remains \((2e+o(1))D\), which exceeds \(2D\) for sufficiently large \(D\).
      Thus the same certificates and asymptotic size guarantee hold with the recomputed parameters.
      For a density bound \(b\) supplied as a rational of polynomial bit length, this yields expected polynomial bit complexity, with a polynomial exponent independent of \(t\) in \cref{rem:ktt-growing-t}.
\end{proof}

\section{Occupancy comparison and conditional mixing}
\label{sec:mixing-comparison}

The neighborhood update condition we use for \cref{thm:transfer} concerns the expected value of the occupancy score.
The following example distinguishes it from convergence of the full conditional probability distribution: it rules out approximate conditional sampling by any prescribed update sequence of polynomial length at the fugacity of \cref{cor:ktt-glauber} with \(b=0\) and \(t=2\) (i.e.\ the case of \(K_{1,2,2}\)-free graphs).

We first define the mixing time and modified log-Sobolev constant used below.
Let \(K\) be an irreducible, aperiodic Markov kernel on a finite state space \(\Omega\), with stationary distribution \(\pi\).
For \(0<\epsilon<1\), the worst-case mixing time is \[ t_{\mathrm{mix}}(K,\epsilon) =\min\left\{j\in\ZZ_{\geq0}:\max_{x\in\Omega} \tv\bigl(K^j(x,\cdot),\pi\bigr)\leq\epsilon\right\}.
\]
When \(K\) is reversible with respect to \(\pi\), define the entropy and Dirichlet form by
\begin{align*}
      \operatorname{Ent}_\pi(f)
       & =\E_\pi[f\log f]-\E_\pi[f]\log\E_\pi[f], \\
      \mathcal E_K(f,h)
       & =\E_\pi[f(h-Kh)],
\end{align*}
where \((Kh)(x)=\sum_{y\in\Omega}
K(x,y)h(y)\).
The modified log-Sobolev constant \(\alpha_{\mathrm{MLSI}}(K)\) is the largest \(\alpha\geq0\) such that
\begin{equation}\label{eq:mlsi-definition}
      \alpha\operatorname{Ent}_\pi(f)\leq\mathcal E_K(f,\log f)
      \qquad\text{for every }f:\Omega\to(0,\infty).
\end{equation}

\begin{proposition}
      \label{prop:slow-neighborhood}
      There is a family of \(K_{1,2,2}\)-free graphs \(G_q\) on \(\Delta_q+1\) vertices with maximum degree \(\Delta_q\to\infty\) such that, at \[ \lambda_q=(\log\Delta_q)^{-3}, \] the three-sweep averaged certificate \eqref{eq:averaged-word-certificate} holds from every initial configuration with \(\eta=0\), \(\kappa\leq3\Delta_q\), and the parameters of \cref{cor:ktt-glauber} with \(t=2\) and \(b=0\).

      Nevertheless, some feasible exterior conditioning leaves an available neighborhood graph \(F_q\) with conditional law \(\mu_q=\mu_{F_q,\lambda_q}\) and the following property.
      For an absolute constant \(c_0>0\), every prescribed word \(S\) of at most \(\ell\) single-site heat-bath updates on \(F_q\) satisfies
      \begin{equation}\label{eq:separation-word-tv}
            \sup_{I\in\cI(F_q)}\tv\bigl(S(I,\cdot),\mu_q\bigr)
            \geq\frac12-\ell\exp(-c_0\lambda_q\Delta_q).
      \end{equation}
      The same bound holds for any mixture of such words chosen independently of the input configuration and all heat-bath randomness, including mixtures of prefixes.
      In particular, uniform total variation error at most \(1/4\) requires \(\ell\geq\tfrac14\exp(c_0\lambda_q\Delta_q)\).
      The conditional random-scan Glauber chain consequently satisfies
      \begin{equation}\label{eq:separation-mixing-mlsi}
            t_{\mathrm{mix}}(1/4)\geq\exp(\Omega(\lambda_q\Delta_q)),
            \qquad
            \alpha_{\mathrm{MLSI}}\leq\exp(-\Omega(\lambda_q\Delta_q)),
      \end{equation}
      with total update rate one, as above.
\end{proposition}

\begin{proof}
      Let \(q\) tend to infinity through prime powers, and let \(F_q\) be the point--line incidence graph of the projective plane \(\operatorname{PG}(2,q)\).
      This is a bipartite graph with two parts \(L,R\), each with \(m=q^2+q+1\) vertices, and every vertex has degree \(q+1\).
      Two distinct points lie on exactly one common line, so \(F_q\) is \(K_{2,2}\)-free.
      Form \(G_q\) by adjoining a single apex vertex \(u\) which is adjacent to every vertex of \(F_q\).
      Then \(\Delta_q=2m\), the neighborhood of \(u\) induces \(F_q\), and every other neighborhood induces a star.
      Thus every neighborhood is \(K_{2,2}\)-free, and hence \(G_q\) is \(K_{1,2,2}\)-free.
      \Cref{cor:ktt-glauber} with \(t=2\) and \(b=0\) supplies the three-sweep certificate at \(\lambda_q\).
      Pinning \(u\) to be unoccupied leaves precisely \(\mu_q\) on its neighborhood.

      Abbreviate \(\lambda=\lambda_q\), \(\Delta=\Delta_q\), and \(\mu=\mu_q\).
      We first bound the weight of the balanced independent sets \[ \mathcal B=\{I\in\cI(F_q):|I\cap L|=|I\cap R|\}.
      \]
      A nonincident set of \(k\) points and \(k\) lines satisfies
      \(k\leq(q+1)\sqrt q-q<q^{3/2}\) by the projective-plane nonincidence bound~\cite{Sti13}.
      Write \(w_\lambda(\mathcal A)=\sum_{I\in\mathcal A}\lambda^{|I|}\) for a family \(\mathcal A\) of independent sets.
      Since \(0<\lambda\leq1\),
      \begin{equation}\label{eq:separation-balanced-weight}
            w_\lambda(\mathcal B)
            \leq\sum_{k=0}^{\lfloor q^{3/2}\rfloor}\binom mk^2\lambda^{2k}
            \leq\exp\bigl(O(q^{3/2}\log q)\bigr).
      \end{equation}
      Let \(\mathcal A_+\) and \(\mathcal A_-\) be the two strict phases, defined by whether \(|I\cap L|-|I\cap R|\) is positive or negative.
      Each phase contains every nonempty subset of its corresponding bipartition class, so \[ w_\lambda(\mathcal A_\pm)\geq(1+\lambda)^m-1.
      \]
      Choose one phase \(\mathcal A\) with \(a:=\mu(\mathcal A)\leq1/2\), and write \(r=\mu(\mathcal B)/a\).
      Since \(\lambda m=\Theta(q^2/(\log q)^3)\), we have \(q^{3/2}\log q=o(\lambda m)\), and the preceding estimates give, for sufficiently large \(q\),
      \begin{equation}\label{eq:separation-bottleneck}
            r=\frac{w_\lambda(\mathcal B)}{w_\lambda(\mathcal A)}
            \leq\frac{\exp(O(q^{3/2}\log q))}{(1+\lambda)^m-1}
            \leq\exp(-c_0\lambda\Delta).
      \end{equation}
      This is the classic balanced-phase bottleneck used in slow mixing arguments for the hard-core model on bipartite graphs; see e.g.~\cite[Sec.~2]{GT06}.

      The bottleneck obstructs every prescribed update word.
      Let \(R_v\) be the single-site heat-bath kernel at \(v\in V(F_q)\), and start from \(\nu_0=\mu(\cdot\mid\mathcal A)\).
      Each \(R_v\) leaves \(\mu\) invariant, so the pointwise domination \(\nu_0\leq\mu/a\) persists after any prescribed sequence: \[ \nu_j=\nu_0R_{v_1}\cdots R_{v_j}\leq\mu/a, \qquad \nu_j(\mathcal B)\leq r.
      \]
      A single-site update changes \(|I\cap L|-|I\cap R|\) by at most one.
      Thus a trajectory starting in \(\mathcal A\) must visit \(\mathcal B\) before leaving that phase.
      The union bound gives \(\nu_\ell(\mathcal A)\geq1-\ell r\), and hence \[ \tv(\nu_\ell,\mu) \geq\nu_\ell(\mathcal A)-\mu(\mathcal A) \geq1-a-\ell r\geq\frac12-\ell r.
      \]
      Averaging the same event bound over any configuration-independent mixture of words of length at most \(\ell\) preserves it.
      Since \(\nu_0\) is a mixture of deterministic initial configurations, convexity of total variation gives \eqref{eq:separation-word-tv} in both cases.
      In particular, some initial configuration has the displayed error for each choice of word or mixture.

      Random-scan dynamics is such a mixture of words, proving the mixing-time bound in \eqref{eq:separation-mixing-mlsi}.
      For the modified log-Sobolev bound, let \(\widehat K=(2m)^{-1}\sum_vR_v\) and let \(g\) be its spectral gap.
      The only transitions from \(\mathcal A\) to its complement enter \(\mathcal B\), so reversibility bounds their stationary flow by \(\mu(\mathcal B)\).
      Testing the spectral gap with \(\indicator{\mathcal A}\) therefore gives \[ g\leq\frac{\sum_{I\in\mathcal A,\,J\notin\mathcal A} \mu(I)\widehat K(I,J)}{a(1-a)}\leq2r.
      \]
      Linearizing \eqref{eq:mlsi-definition} at \(f=1\) gives \(\alpha_{\mathrm{MLSI}}\leq2g\leq4r\), which proves the remaining bound.
\end{proof}

Thus polynomial-length occupancy certificates require strictly less than uniform conditional neighborhood sampling, even at the fugacity of our \(K_{1,2,2}\)-free independent-set guarantee.
The incidence neighborhoods have degree \(q+1\), so this construction does not settle conditional mixing when the local maximum average degree is fixed, a special case of \cref{cor:asymptotic}.

\appendix
\crefalias{section}{appendix}
\section{Chromatic local occupancy and correspondence coloring}
\label{app:chromatic-occupancy}

A slight strengthening of the local occupancy condition~\eqref{eq:intro-occupancy} has been used to bound the (correspondence) chromatic number~\cite{DKPS20a,DKPS20b}.
The strengthening replaces the expectation \(E_F(\lambda)\) by a quantile estimate, and it also asserts that for every \(F\) at least one of the two terms is at least one, rather than their sum.
In many applications this extra strength can be obtained without leading-order losses in the quantity \(\beta + \Delta\gamma\).
Keep the notation \(p=\lambda/(1+\lambda)\), and for a separate probability parameter \(q\in(0,1)\) write \[ Q_{F,q}(\lambda)=\max\bigl\{a\in\ZZ_{\geq0}: \Pr_{\mu_{F,\lambda}}(|I|\geq a)\geq q\bigr\}.
\]
Following~\cite{DKPS20a}, we say that the hard-core model at fugacity \(\lambda\) on \(G\) has \emph{\(q\)-chromatic local occupancy} with parameters \((\beta,\gamma)\) if, for every vertex \(u\) and every subgraph \(F\subseteq G[N(u)]\),
\begin{equation}\label{eq:chromatic-lococc}
      \frac{\beta p}{Z_F(\lambda)}\geq1
      \qquad\text{or}\qquad
      \gamma q Q_{F,q}(\lambda)\geq1.
\end{equation}
Here, \(F\) need not be induced which is important for the application to correspondence coloring.
Our hereditary density condition \eqref{eq:ktt-hereditary-budget} also holds for arbitrary subgraphs because deleting edges can only decrease the number of copies of \(K_{t,t}\).

Correspondence coloring is a generalization of list coloring and ordinary graph coloring, defined as follows.
A correspondence assignment gives each vertex \(u\) a finite list \(L(u)\) and each edge \(uv\) a matching between \(L(u)\) and \(L(v)\) specifying forbidden pairs of colors.
A coloring consists of giving each vertex one color from its list without choosing a forbidden pair on any edge.
The correspondence chromatic number \(\chi_c(G)\) is the least integer \(k\) such that every such assignment with lists of size \(k\) admits a coloring.
This generalizes list coloring and ordinary coloring, so \(\chi(G)\leq\chi_\ell(G)\leq\chi_c(G)\); we defer to~\cite{DKPS20a} for further details.

We use the following specialization of the main correspondence coloring theorem of~\cite{DKPS20a}.
We restate this specialization in our notation for convenience, and refer to~\cite{DKPS20a} for the general statement and proof.

\begin{theorem}[{Special case of~\cite[Thm.~13]{DKPS20a}}]\label{thm:chromatic-lococc}
      Let \(G\) have maximum degree at most \(\Delta\geq2\), and suppose it satisfies \(q\)-chromatic local occupancy with parameters \((\lambda,\beta,\gamma)\), for some \(q\in(0,1)\).
      Write \(p=\lambda/(1+\lambda)\).
      Suppose \(0<\iota<1\) and \(\ell>0\) satisfy, for every vertex \(u\in V(G)\),
      \begin{equation}\label{eq:chromatic-transfer-concentration}
            \ell\geq\frac2{\iota^2}\log(16\Delta^3),
            \qquad
            d_G(u)\gamma\geq\frac2{\iota^2q}\log(16\Delta^3),
      \end{equation}
      and, for every subgraph \(F\) of a neighborhood in \(G\),
      \begin{equation}\label{eq:chromatic-transfer-partition}
            |V(F)|\geq\ell/8\quad\Longrightarrow\quad
            Z_F(\lambda)\geq8\Delta^4.
      \end{equation}
      Then \[ \chi_c(G)\leq \left\lceil\frac{p\ell\beta+\gamma\Delta}{1-\iota}\right\rceil.
      \]
\end{theorem}

We first establish the stronger chromatic local occupancy condition in our setting.
To do this, we strengthen the estimates from \cref{prop:ktt-occupancy} to lower-tail bounds for the size of a hard-core independent set and apply the correspondence coloring framework of~\cite{DKPS20a}.
Throughout this appendix, the integer \(t\geq1\) and hereditary density bound \(b\geq0\) may both vary with the maximum-degree bound \(\Delta\geq3\).
We occasionally use different notation and parameters from those in the main paper argument.
Write
\begin{align}\label{eq:ktt-chromatic-scales}
      s & =(b+1)^{1/(2t-1)},        &
      L & =\log\Delta,              &
      H & =\log\frac{\Delta}{sL^t}.
\end{align}
We work in the regime
\begin{align}\label{eq:ktt-chromatic-regime}
      H & \to\infty, & t\log H & =o(H).
\end{align}
All asymptotic estimates below are uniform in \(b\) and \(t\), with absolute implicit constants.
In particular, \(H\leq L\), \(t/H^2=o(1)\), and \(H^{2t+2}e^{-H}=\exp(-H+o(H))\to0\).
Define
\begin{align}\label{eq:ktt-chromatic-density-parameters}
      \lambda & =\frac1{sL^{t-1}
                       H^{2t}}, & M & =\frac{4eL}{\lambda}=4esL^tH^{2t}, & c & =\exp\left(-\frac{4t(4e)^{1-1/t}}{H^2}\right).
\end{align}

Choosing \(q\) independently of \(\lambda\) lets us control the required lower quantile uniformly as \(b\) and \(t\) vary.
The additional ingredient beyond \cref{prop:ktt-occupancy} is the following lower-tail estimate.

\begin{lemma}\label{lem:ktt-lower-tail}
      Use the established parameters in \eqref{eq:ktt-chromatic-scales} and \eqref{eq:ktt-chromatic-density-parameters}, and assume that \(H\geq1\) and \(H^2\geq8et/(4e)^{1/t}\).
      Let \(F\) be a graph on \(k\) vertices with \(K_{t,t}\)-density at most \(b\), and let \(X=|I|\) for \(I\sim\mu_{F,\lambda}\).
      If \(k\leq M\), then
      \begin{equation}\label{eq:ktt-small-partition}
            \log Z_F(\lambda)\geq c\lambda k,
      \end{equation}
      and for every \(0<\eta<1\),
      \begin{equation}\label{eq:ktt-small-tail}
            \Pr\bigl(X\leq(1-\eta)c\lambda k\bigr)
            \leq\exp(-\eta^2c\lambda k/2).
      \end{equation}
      If \(k>M\), then
      \begin{equation}\label{eq:ktt-large-tail}
            \Pr(X\leq L)\leq2^{-L}.
      \end{equation}
\end{lemma}

\begin{proof}
      The assertions for \(k=0\) are immediate, so suppose \(k\geq1\), and let \(d\) be the average degree of \(F\).
      If \(k\leq M\), then \(M\geq s\), so the average degree bound~\eqref{eq:ktt-degree} offered by \cref{lem:ktt-weighted} gives \[ \lambda(d+1)\leq4t\lambda s(M/s)^{1-1/t} =\frac{4t(4e)^{1-1/t}}{H^2}.
      \]
      Thus, the elementary estimate~\eqref{eq:ktt-elementary} holds throughout the interval below \(\lambda\) in the sense that for fugacity \(0< x\le\lambda\) we have
      \begin{equation}\label{eq:ktt-small-interval}
            E_F(x)\geq kx e^{-(d+1)x}\geq ckx.
      \end{equation}
      Integrating \(E_F(x)/x=(\log Z_F)'(x)\) from \(0\) to \(\lambda\) then gives~\eqref{eq:ktt-small-partition}.
      A bound on the mean at fugacity \(\lambda\) alone does not control the lower tail, and hard-core occupancy indicators need not be negatively correlated.
      We therefore use~\eqref{eq:ktt-small-interval} at all smaller fugacities to bound the Laplace transform: for every \(\theta>0\),
      \begin{align*}
            \E e^{-\theta X}
             & =\frac{Z_F(\lambda e^{-\theta})}{Z_F(\lambda)} \\
             & =\exp\left(-\int_{\lambda e^{-\theta}}^\lambda
            \frac{E_F(x)}x\,dx\right)
            \leq\exp\bigl(-c\lambda k(1-e^{-\theta})\bigr).
      \end{align*}
      Choose \(\theta=-\log(1-\eta)\) and apply Markov's inequality to get \[ \Pr\bigl(X\leq(1-\eta)c\lambda k\bigr) \leq\exp\bigl(-c\lambda k[\eta+(1-\eta)\log(1-\eta)]\bigr) \leq\exp(-\eta^2c\lambda k/2), \] proving \eqref{eq:ktt-small-tail}.

      If \(k>M\), apply the estimate~\eqref{eq:ktt-coarse} obtained in the proof of \cref{prop:ktt-occupancy} at fugacity \(\lambda/2\), together with the average degree bound~\eqref{eq:ktt-degree}, to obtain \[ E_F(\lambda/2) \geq\frac1e\min\left\{\frac{\lambda k}2,\frac{k}{d+1}\right\} \geq\frac1e\min\left\{2eL, \frac{(4e)^{1/t}LH^2}{4t}\right\} \geq2L, \] where the last inequality uses \(H^2\geq8et/(4e)^{1/t}\).
      Since \(E_F\) is increasing, Markov's inequality now gives the required tail bound \[ \Pr(X\leq L) \leq2^L\E 2^{-X} =2^L\frac{Z_F(\lambda/2)}{Z_F(\lambda)} \leq\exp\left(L\log2-\int_{\lambda/2}^\lambda \frac{2L}x\,dx\right) =2^{-L}.
            \qedhere
      \]
\end{proof}

\begin{proposition}
      \label{prop:ktt-chromatic-occupancy}
      Let \(t\geq1\) be an integer, let \(b\geq0\), and suppose \(G\) has maximum degree at most \(\Delta\geq3\) and hereditary \(K_{t,t}\)-density at most \(b\) in neighborhoods.
      Use the parameters in \eqref{eq:ktt-chromatic-scales} and \eqref{eq:ktt-chromatic-density-parameters}, in the regime \eqref{eq:ktt-chromatic-regime}.
      Write \(p=\lambda/(1+\lambda)\), and set
      \begin{equation}\label{eq:ktt-chromatic-threshold}
            q=1-\frac1H,\qquad
            z_0=H-(2t+3)\log H,\qquad
            \epsilon=\sqrt{\frac{8\log H}{H}},\qquad
            a=\left\lfloor(1-\epsilon)cz_0\right\rfloor.
      \end{equation}
      For sufficiently large \(H\), we have \(a\geq1\), and the hard-core model on \(G\) at fugacity \(\lambda\) has \(q\)-chromatic local \((\beta,\gamma)\)-occupancy for
      \begin{equation}\label{eq:ktt-chromatic-parameters}
            \beta=\frac{e^{z_0}}p
            =\frac{(1+\lambda)\Delta}{LH^3},
            \qquad
            \gamma=\frac1{qa}
            =\left(1+O\!\left(\sqrt{\frac{\log H}{H}}+\frac{t\log H}{H}\right)\right)
            \frac1H.
      \end{equation}
      The implicit constant is absolute.
\end{proposition}

\begin{proof}
      Under \eqref{eq:ktt-chromatic-regime}, we have \(z_0/H=1-O(t\log H/H)\), \(\epsilon=o(1)\), and \(c=1-O(t/H^2)\), with absolute implicit constants.
      Thus \(a\geq1\) for sufficiently large \(H\), and the finite condition in \cref{lem:ktt-lower-tail} also holds.
      Fix a subgraph \(F\) of a neighborhood, and write \(k=|V(F)|\).
      If \(Z_F(\lambda)\leq e^{z_0}\), then the first alternative in \eqref{eq:chromatic-lococc} holds.
      This includes the empty graph.
      Suppose instead that \(Z_F(\lambda)>e^{z_0}\).

      If \(k\leq M\), then \(\lambda k\geq\log Z_F(\lambda)>z_0\) by \eqref{eq:ktt-elementary}.
      \Cref{lem:ktt-lower-tail} and \(a\leq(1-\epsilon)cz_0\) therefore give \[ \Pr_{\mu_{F,\lambda}}(|I|<a) \leq\exp(-\epsilon^2c\lambda k/2) \leq\exp(-\epsilon^2cz_0/2) \leq H^{-2}\leq1-q.
      \]
      The penultimate inequality holds once \(cz_0\geq H/2\).
      If \(k>M\), then \(a\leq H\leq L\), and \eqref{eq:ktt-large-tail} gives \[ \Pr_{\mu_{F,\lambda}}(|I|<a)\leq2^{-L}\leq H^{-1}=1-q.
      \]
      Thus, in both cases \(Q_{F,q}(\lambda)\geq a\), proving the second alternative in \eqref{eq:chromatic-lococc}.
      Finally, \(a/H=1-O(\epsilon+t\log H/H)\), including the error from rounding down, and \(q=1-H^{-1}\), which give the estimate for \(\gamma\).
      The identity for \(\beta\) follows from \(e^H=\Delta/(sL^t)\).
\end{proof}

Having established chromatic local occupancy, we can apply the correspondence coloring theorem of~\cite{DKPS20a} to obtain a correspondence coloring bound with leading constant one.

Applying \cref{thm:chromatic-lococc} with the parameters from \cref{prop:ktt-chromatic-occupancy} gives the following uniform bound.

\begin{theorem}
      \label{thm:ktt-correspondence}
      Let \(t\geq1\) be an integer and let \(b\geq0\), both allowed to vary with \(\Delta\).
      Let \(G\) have maximum degree at most \(\Delta\geq3\) and hereditary \(K_{t,t}\)-density at most \(b\) in neighborhoods.
      Write \[ H=\log\frac{\Delta}{(b+1)^{1/(2t-1)}(\log\Delta)^t}.
      \]
      If \(H\to\infty\) and \(t\log H=o(H)\), then
      \[
            \chi_c(G)\leq
            \left(1+O\!\left(\sqrt{\frac{\log H}{H}}+\frac{t\log H}{H}\right)\right)
            \frac{\Delta}{H}.
      \]
      The implicit constant is absolute, uniformly in \(b\) and \(t\).
      In particular, if \(b\) is fixed and \(t=o(\log\Delta/\log\log\Delta)\), or if \(t\) is fixed and \(b+1 = \Delta^{o(1)}\), then \(\chi_c(G)\leq(1+o(1))\Delta/\log\Delta\).
\end{theorem}

\begin{proof}
      Use the parameters from \cref{prop:ktt-chromatic-occupancy}, and set
      \begin{align}
            \ell  & =8M=32esL^tH^{2t},                                               &
            \iota & =\sqrt{\frac{7L}{\ell}},                                         &
            K     & =\left\lceil\frac{p\ell\beta+\gamma\Delta}{1-\iota}\right\rceil.
      \end{align}
      The contribution of \(\beta\) to the bound \(K\) on correspondence chromatic number given by \cref{thm:chromatic-lococc} is roughly
      \begin{equation}\label{eq:ktt-chromatic-beta-cost}
            p\ell\beta=\ell e^{z_0}=\frac{32e\Delta}{H^3}.
      \end{equation}
      Since \(\iota=O(H^{-1})\) and \(\Delta\geq e^H\), conclusion~\eqref{eq:ktt-chromatic-parameters} of \cref{prop:ktt-chromatic-occupancy} gives the required asymptotics of \(K\).

      To justify the application of \cref{thm:chromatic-lococc}, we first verify \eqref{eq:chromatic-transfer-partition}.
      If a neighborhood subgraph \(F\) has at least \(\ell/8=M\) vertices, choose an induced subgraph \(J\) of \(F\) on \(\lfloor M\rfloor\) vertices.
      The \(K_{t,t}\)-density bound of \(b\) holds for \(J\), so \eqref{eq:ktt-small-partition} from \cref{lem:ktt-lower-tail} gives \[ \log Z_F(\lambda)\geq\log Z_J(\lambda) \geq c\lambda\lfloor M\rfloor \geq4ecL-c\lambda>\log(8\Delta^4).
      \]
      The last inequality holds for sufficiently large \(H\), uniformly in \(b\) and \(t\), because \(c=1-O(t/H^2)=1-o(1)\), \(\lambda\leq H^{-2}\), and \(L\geq H\).
      Moreover, \(2\log(16\Delta^3)\leq7L\) for large \(H\), and hence \[ \frac2{\iota^2}\log(16\Delta^3)\leq\ell, \] proving the first condition in \eqref{eq:chromatic-transfer-concentration}.

      To ensure the degree condition of \cref{thm:chromatic-lococc}, repeatedly delete any vertex whose current degree is less than \[ d_0=\frac2{\iota^2q\gamma}\log(16\Delta^3) =\frac{2a}{\iota^2}\log(16\Delta^3)\leq a\ell\leq H\ell.
      \]
      Let \(G_*\) be the remaining graph.
      The hereditary \(K_{t,t}\)-density condition in neighborhoods, chromatic local occupancy, and the partition function condition all pass to \(G_*\), with the same parameters and the same upper bound \(\Delta\) on maximum degree.
      If \(G_*\) is nonempty, it satisfies both conditions in \eqref{eq:chromatic-transfer-concentration}, so \cref{thm:chromatic-lococc} colors it from any correspondence assignment with lists of size at least \(K\).
      If \(G_*\) is empty, start with the empty coloring instead.
      Since \(K\geq\Delta/H\) and \[ \frac{d_0}{K}\leq\frac{H^2\ell}{\Delta} =32eH^{2t+2}e^{-H}=\exp(-H+o(H))\to0, \] restore the deleted vertices in reverse order, choosing a color greedily at each step.
      Each vertex then has fewer than \(d_0<K\) already colored neighbors, and each such neighbor forbids at most one color by the matching condition.
      Hence every correspondence assignment with lists of size at least \(K\) admits a coloring of \(G\).
\end{proof}

The same argument also gives a bound assuming only a total number of copies in each neighborhood, which facilitates comparison with earlier bounds from~\cite{ADK26}.

\begin{corollary}\label{cor:ktt-copy-correspondence}
      Let \(G\) have maximum degree at most \(\Delta\geq3\).
      Let \(t\geq1\) be an integer, \(B\geq0\), and allow both to vary with \(\Delta\).
      Suppose every neighborhood of \(G\) contains at most \(B\) copies of \(K_{t,t}\), and write \[ D_B=\log\frac{\Delta}{(B+1)^{1/(2t)}},\qquad H_B=D_B-t\log\log\Delta.
      \]
      If \(H_B\to\infty\) and \(t\log H_B=o(H_B)\), then
      \[
            \chi_c(G)\leq
            \left(1+O\!\left(\sqrt{\frac{\log H_B}{H_B}}+\frac{t\log H_B}{H_B}\right)\right)
            \frac{\Delta}{H_B}.
      \]
      The implicit constant is absolute, uniformly in \(B\) and \(t\).
      In particular, if \(D_B/(t\log\log\Delta)\to\infty\), then \(\chi_c(G)\leq(1+o(1))\Delta/D_B\).
      This includes \(t=o(\log\Delta/\log\log\Delta)\) and \(B+1\leq\Delta^{2t(1-\delta)}\) for any fixed \(0<\delta<1\).
\end{corollary}

\begin{proof}
      Every subgraph \(F\) of a neighborhood on \(k\geq1\) vertices contains at most \(\min\{B,k^{2t}\}\) copies of \(K_{t,t}\).
      Consequently its density is at most \[ \frac{2t}{k}\min\{B,k^{2t}\} \leq2tB^{1-1/(2t)}.
      \]
      Apply \cref{thm:ktt-correspondence} with
      \(b=2tB^{1-1/(2t)}\).
      For this choice, \[ 0\leq\log (b+1)^{1/(2t-1)} -\frac1{2t}\log(B+1) \leq\frac{\log(2t+1)}{2t-1}\leq\log3.
      \]
      Hence \(H=H_B+O(1)\) uniformly in \(t,B\), so the growth conditions and the displayed error estimate transfer from \(H_B\) to \(H\).
      The remaining assertions follow because \(D_B/(t\log\log\Delta)\to\infty\) implies \(H_B=(1-o(1))D_B\to\infty\) and \(t\log H_B\leq t\log\log\Delta=o(H_B)\).
\end{proof}

\subsection{Comparison with earlier coloring bounds}
\label{subsec:coloring-comparison}

For fixed \(t\) and \(b+1=\Delta^{o(1)}\), \eqref{eq:ktt-degree} bounds the maximum average degree in every neighborhood by \(\Delta^{1-1/t+o(1)}\).
Applying the earlier local-occupancy analysis for bounded neighborhood maximum average degree from~\cite{DKPS20a} therefore gives \(\chi_c(G)\leq(t+o(1))\Delta/\log\Delta\).
This is an application of that method to the hereditary density hypothesis; in this work we improve the leading constant to one independent of \(t\).

We also improve the result of~\cite{ABD25} by obtaining a leading constant of one in the same range of \(t\ge 4\) for which their nibble method gives a leading constant of four.

For comparison under a hypothesis that bounds the total number of copies of \(K_{t,t}\) in each neighborhood, keep \(t\) fixed and set \(B_* = \max\{B,1/2\}\).
For every fixed \(\epsilon>0\) and sufficiently large \(\Delta\), \cite[Cor.~15]{ADK26} specialized to \(K_{t,t}\) assumes
\(B_*\leq\Delta^{2t/10}\) and gives
\[
      \chi_c(G)\leq C\frac{\Delta}{\log(\Delta B_*^{-1/(2t)})},
      \qquad
      C=\begin{cases}
            4+\epsilon, & B_*\leq\Delta^{2t\epsilon/200}, \\
            8,          & \text{otherwise}.
      \end{cases}
\]
Since \(\log(B+1)=\log B_*+O(1)\), our \cref{cor:ktt-copy-correspondence} has the same asymptotic denominator throughout this range, with leading constant \(1+o(1)\).

\section{AI methodology}

The direction and high-level proof ideas in this work were developed by the authors.
We used GPT-6 Astra to assist with basic analytic calculations and to check our main findings.
The tool materially affected the proofs of \cref{lem:ktt-weighted,lem:ktt-lower-tail}, and \cref{prop:ktt-occupancy,prop:ktt-chromatic-occupancy}.
These results are standard applications of ideas from the literature, including the K\H{o}v\'ari--S\'os--Tur\'an Theorem~\cite{KST54}, and analyzing lower tails through exponential generating functions in the manner of standard proofs of the Chernoff bound.
The tool also assisted with the sweep analysis, \cref{lem:ktt-sweep-mean}.
The construction in \cref{prop:slow-neighborhood} was developed by the tool, though the high-level idea of a balanced-phase bottleneck in a balanced bipartite graph is intuitive and present in the literature, e.g.~\cite{GT06}.
The specific use of the projective plane in \cref{prop:slow-neighborhood} was suggested by the AI.
The authors verified the correctness and originality of all content, including references.

\printbibliography

\end{document}